\documentclass[twocolumn,prl,10pts,nofootinbib,superscriptaddress]{revtex4-1}
\usepackage[utf8]{inputenc}

\def\mySections#1{{\bf #1.} } 
\usepackage{hyperref}
\usepackage{braket}  
\usepackage[dvips]{graphicx}
\usepackage{hhline}
\usepackage{epsfig,amsmath,amssymb,verbatim,mathrsfs,array,layout,textcomp,amssymb,latexsym,slashed,graphicx,booktabs,color,mathtools}

\newcommand{\beq}{\begin{equation}}% can be used as {equation} or {eqnarray}
\newcommand{\eeq}{\end{equation}}

\def\beqa{\begin{eqnarray}}
\def\eeqa{\end{eqnarray}}
\def\bea{\begin{eqnarray}}
\def\eea{\end{eqnarray}}

\newcommand{\bv}{\left(\begin{array}{c}}
\newcommand{\ev}{\end{array}\right)}
\newcommand{\bmtwo}{\left(\begin{array}{cc}}
\newcommand{\bmthree}{\left(\begin{array}{ccc}}
\newcommand{\emn}{\end{array}\right)}
\newcommand{\bmtwoc}{\left\{\begin{array}{cc}}
\newcommand{\bmthreec}{\left\{\begin{array}{ccc}}
\newcommand{\emnc}{\end{array}\right\}}
\newcommand{\ba}{\begin{array}}
\newcommand{\ea}{\end{array}}
\newcommand{\be}{\begin{eqnarray}}
\newcommand{\ee}{\end{eqnarray}}

\newcommand{\GeV}{\text{ GeV}}
\newcommand{\TeV}{\text{ TeV}}
\newcommand{\MeV}{\text{ MeV}}

\definecolor{readableRTD}{rgb}{0.7,0.1,0.2}

\definecolor{readableMG}{rgb}{0.0,0,0.5}

\def\lsim{\mathrel{\rlap{\lower4pt\hbox{\hskip1pt$\sim$}}
     \raise1pt\hbox{$<$}}}         %less than or approx. symbol
\def\gsim{\mathrel{\rlap{\lower4pt\hbox{\hskip1pt$\sim$}}
     \raise1pt\hbox{$>$}}}         %greater than or approx. symbol

\begin{document}
%\vspace*{-30mm}
%\font\mini=cmr10 at 0.8pt

\title{
Crunching Dilaton, Hidden Naturalness
}

\author{Csaba Cs\'aki}
\affiliation{Department of Physics, LEPP, Cornell University, Ithaca, NY 14853, USA}

\author{Raffaele Tito D'Agnolo}
\affiliation{Institut de Physique Th\'eorique, Universit\'e Paris Saclay, CEA, F-91191 Gif-sur-Yvette, France}

\author{Michael Geller}
\affiliation{School of Physics and Astronomy, Tel-Aviv University, Tel-Aviv 69978, Israel}
\author{Ameen Ismail}
\affiliation{Department of Physics, LEPP, Cornell University, Ithaca, NY 14853, USA}
\begin{abstract}
We introduce a new approach to the Higgs naturalness problem, where the value of the Higgs mass is tied to cosmic stability and the possibility of a large observable Universe. The Higgs mixes with the dilaton of a CFT sector whose true ground state has a large negative vacuum energy. If the Higgs VEV is non-zero and below $\mathcal{O}({\rm TeV})$, the CFT also admits a second metastable vacuum, where the expansion history of the Universe is conventional. 
 As a result, only Hubble patches with unnaturally small values of the Higgs mass support inflation and post-inflationary expansion, while all other patches rapidly crunch. The elementary Higgs VEV driving the dilaton potential is the essence of our new solution to the hierarchy problem.
 The main experimental prediction is a light dilaton field in the 0.1-10 GeV range that mixes with the Higgs. Part of the viable parameter space has already been probed by measurements of rare B-meson decays,  and the rest will be fully explored by future colliders and experiments searching for light, weakly-coupled particles.  
\end{abstract}
\maketitle

%%%%%%%%%%%%%%%%%%%%%%%%%%%%%%%%%%%%%%%%%%%%%%%%%%%%%%%%%%%%%%%%%%%%%%%%%%%
%%%%%%%%%%%%%%%%%%%%%%%%%%%%%%%%%%%%%%%%%%%%%%%%%%%%%%%%%%%%%%%%%%%%%%%%%%%
\section{Introduction}
%%%%%%%%%%%%%%%%%%%%%%%%%%%%%%%%%%%%%%%%%%%%%%%%%%%%%%%%%%%%%%%%%%%%%%%%%%%
%%%%%%%%%%%%%%%%%%%%%%%%%%%%%%%%%%%%%%%%%%%%%%%%%%%%%%%%%%%%%%%%%%%%%%%%%%%

While the discovery of the Higgs boson at the Large Hadron Collider~\cite{Aad:2012tfa,Chatrchyan:2012xdj} marked a triumph for the Standard Model of particle physics, the measured values of its mass and vacuum expectation value (VEV) remain unexplained. The Higgs VEV determines the mass scales of all elementary particles in the Standard Model (SM) and hence greatly influences all phenomenology at the weak scale and below, including, for instance, the stability of nuclei and  stars. 

In the SM the Higgs mass squared $m_H^2$ corresponds to a relevant operator not protected by any symmetries; it is therefore expected to be sensitive to new physics at high scales and receive corrections hierarchically larger than its measured value.  In the absence of new mass scales beyond the SM this would be a moot point, but several physical thresholds exist that require the appearance of new physics, in particular the unification scale at $M_{\rm GUT} \sim 10^{14} m_H$, (which avoids the $U(1)_Y$ Landau pole at higher scales), and ultimately quantum gravity at the Planck scale $M_{\rm Pl}\sim 10^{17} m_H$.

The traditional approaches for explaining the lightness of the Higgs boson relative to these scales predict new colored particles around a TeV. These approaches are under ever-increasing experimental pressure, since the Large Hadron Collider (LHC) did not discover any  
evidence for physics beyond the SM at the TeV scale~\cite{ATLASResults, CMSResults}. It is then natural to wonder whether there are other approaches to the problem that are not tied to the appearance of new particles at a TeV, or more specifically observable ``top partners''. One well-known idea is neutral naturalness. In its classic implementation, twin Higgs~\cite{Chacko:2005pe}, the top partners interact only very weakly with the SM and are not produced at observable rates at colliders. These models, however, generally predict deviations in Higgs coupling measurements at the LHC that have not yet been observed. 

A radical departure from these symmetry-based solutions are mechanisms for cosmological selection of the Higgs mass~\cite{Graham:2015cka, Arkani-Hamed:2016rle, Geller:2018xvz, Cheung:2018xnu, Strumia:2020bdy, Giudice:2019iwl}.
 In these approaches, the corrections to the Higgs potential from high energy scales are unsuppressed; the lightness of the Higgs is instead realized by the cosmological dynamics of new fields. 
   
The most extreme break with standard BSM lore is to rely on a fully anthropic reasoning for the existence of a light Higgs in the context of a multiverse with several patches, each corresponding to a different Higgs mass/VEV. This approach has been advocated by~\cite{Agrawal:1997gf} and is also partly the motivation behind split SUSY models~\cite{ArkaniHamed:2004fb}. 

We introduce a novel approach to the Higgs hierarchy problem that incorporates the most appealing elements from the three classes of solutions described above. 
  The Higgs VEV plays a dominant role in the existence of a vacuum where the cosmological constant (CC) can be positive, allowing for inflation and post-inflationary expansion of the universe.  To this end, we introduce a spontaneously broken CFT which has a large and negative vacuum energy in its true ground state leading to an immediate cosmic crunch. This sector is charged under the EW gauge group and couples to the fundamental Higgs field. If the Higgs VEV is sufficiently small---but not zero---the mixing of the Higgs with the dilaton field (setting the scale for the breaking of scale invariance) can generate a second, cosmologically long-lived vacuum with a small positive CC. The selection of a small Higgs VEV is thus dynamical and happens during the early history of the Universe. For this to be possible we need Higgs VEVs close to the weak scale to have an $\mathcal{O}(1)$ effect on the dilaton potential. This is naturally realized as a consequence of scale invariance, like in symmetry-based solutions to the hierarchy problem.

As in the anthropic approach, we assume the presence of a multiverse with a large number of causally disconnected Hubble patches, each having a different Higgs mass parameter.  Only patches where the Higgs VEV is ${\cal O}(100\ {\rm GeV})$ survive crunching and undergo inflation, leading to a dynamical selection of regions with small Higgs mass/VEV that appear unnatural to a low energy observer. In contrast to the regular anthropic approach where the Higgs is typically much heavier in most of the universe, here electroweak (EW)-scale masses dominate the universe on cosmological time scales. 

Our mechanism is fully testable, making several experimental predictions for current and future colliders. First, we expect KK partners of the Higgs and electroweak gauge bosons to lie close to the EW scale, but we do not predict any light top partners. In fact, our EW KK states have nothing to do with the cancellation of quadratic contributions to the Higgs mass. Current LHC bounds on such states stretch to the TeV range, generating a ``little hierarchy"  in our model. 

Second, due to this little hierarchy, the dilaton must be light, below $\sim 10$ GeV---testable in rare $B$ decays, searches for weakly-coupled light particles, future precision $Z$ experiments, and heavy-ion collisions. UV completions of the model that can generate the little hierarchy without fine-tuning  will likely require a form of supersymmetry that results in a split SUSY-like spectrum.

%%%%%%%%%%%%%%%%%%%%%%%%%%%%%%%%%%%%%%%%%%%%%%%%%%%%%%%%%%%%%%%%%%%%%%%%%%%
%%%%%%%%%%%%%%%%%%%%%%%%%%%%%%%%%%%%%%%%%%%%%%%%%%%%%%%%%%%%%%%%%%%%%%%%%%%
\section{The Basic Concept}
%%%%%%%%%%%%%%%%%%%%%%%%%%%%%%%%%%%%%%%%%%%%%%%%%%%%%%%%%%%%%%%%%%%%%%%%%%%
%%%%%%%%%%%%%%%%%%%%%%%%%%%%%%%%%%%%%%%%%%%%%%%%%%%%%%%%%%%%%%%%%%%%%%%%%%%

We assume a landscape of Higgs mass values with a cutoff at the scale $\Lambda$: 
\begin{equation}
V_H(H) = -m^2_{H,i} H^\dagger H +\lambda (H^\dagger H)^2   
\label{eq:Hpotential}
\end{equation}
where a typical $m_{H,i}^2$ is ${\cal O} (\Lambda^2)$. We remain agnostic as to how this landscape is generated and populated.  

We introduce dynamics which can support the expansion of the universe only when the Higgs VEV, $h\equiv \langle H^0 \rangle$, is in a finite range, 
\be
H_I \lesssim h_{\rm min} \lesssim h \leq h_{\rm crit} \simeq {\cal O}(1~{\rm TeV})\, ,
\label{eq:hrange}
\ee
and cause an immediate crunch for other values. In the above equation $H_I$ is Hubble during inflation. Such dynamics excludes all positive and large negative mass terms for the Higgs, and only values of the VEV below the weak scale survive inflation. The mechanism is not sensitive to the minimal value $h_{\rm min}$, which can be generated in many ways briefly discussed in the Cosmological Constraints Section.

The dynamics needed to achieve this is based on the mixing of the Higgs with a spontaneously broken CFT---or a bulk Higgs in the AdS picture with the Higgs potential, Eq.~\eqref{eq:Hpotential}, on the UV brane. The CFT is spontaneously broken via the Goldberger-Wise (GW) mechanism, with the GW minimum for the dilaton $\langle \chi \rangle = \chi_{GW}$ above $\Lambda$ and the scale of inflation $M_{I} \simeq \sqrt{M_{\rm Pl} H_I}$, so that the total vacuum energy in this minimum is always negative, even during the slow-roll regime of inflation. Any patch in which the dilaton has reached this minimum crunches. The heart of our mechanism is the generation of a second minimum for the dilaton by the bulk Higgs VEV, for which the vacuum energy is subdominant to the inflaton vacuum energy; any patch in this second minimum then goes through inflation without crunching. This minimum only exists for a finite range of small Higgs VEVs, set by the parameters of the bulk Higgs. Therefore, only this range of VEVs survive after inflation and until today. These small values are not typical in the landscape, thus generating a hierarchy and an apparent naturalness problem. We assume that one of the usual mechanisms (such as scanning in the multiverse plus anthropic selection) ensures a small positive CC in the shallower metastable minimum, while it cannot overcome the large negative energy of the true minimum.

%%%%%%%%%%%%%%%%%%%%%%%%%%%%%%%%%%%%%%%%%%%%%%%%%%%%%%%%%%%%%%%%%%%%%%%%%%%
%%%%%%%%%%%%%%%%%%%%%%%%%%%%%%%%%%%%%%%%%%%%%%%%%%%%%%%%%%%%%%%%%%%%%%%%%%%
\section{RS model}
%%%%%%%%%%%%%%%%%%%%%%%%%%%%%%%%%%%%%%%%%%%%%%%%%%%%%%%%%%%%%%%%%%%%%%%%%%%
%%%%%%%%%%%%%%%%%%%%%%%%%%%%%%%%%%%%%%%%%%%%%%%%%%%%%%%%%%%%%%%%%%%%%%%%%%%
 
We use the 5D warped description~\cite{Randall:1999ee} of the CFT with the Higgs field in the bulk 
 of AdS space
\begin{equation}
ds^2 = \left(\frac{R}{z}\right)^2 ( \eta_{\mu\nu}dx^\mu dx^\nu-dz^2)\ .
\end{equation}
Here $R=1/k$ is the AdS curvature and the location of the UV brane, while $R'$ is the location of the IR brane, with $\chi = 1/R'$ identified with the dilaton/radion field~\cite{Csaki:1999mp,Csaki:2000zn,Rattazzi:2000hs}. Note that the dilaton defined this way is not canonically normalized---its kinetic term is
\begin{equation}
    \frac{3(N^2-1)}{4\pi^2} (\partial_\mu \chi)^2
\end{equation}
where $N$ is the number of colors in the dual CFT picture, related to the 5D parameters by $N^2-1=16\pi^2 (M_* R)^3$, where $M_*$ is the 5D Planck scale.

The GW stabilization field~\cite{Goldberger:1999uk} gives rise to the usual GW potential for the dilaton~\cite{Rattazzi:2000hs} that we parametrize as 
\begin{equation}
   V_{\rm GW}(\chi) = -\lambda \chi^{4} + \lambda_{\rm GW} \frac{\chi^{4+\delta}}{k^\delta}\, .
\label{eq:GW}
\end{equation}
The $\chi^4$ term is  scale-invariant and in the RS framework can be understood as the effect of some mistuning of the tension of the IR brane and the bulk CC. The $\chi^{4+\delta}$ term is the effect of the small explicit breaking~\footnote{The dilaton is a non-compact Goldstone boson, thereby evading the type of generic problems a shift symmetric relaxion field poses~\cite{Gupta:2015uea}, while its potential can still be controlled by the amount of explicit breaking of scale invariance, except for the quartic term which is fully scale invariant.} of scale invariance by an operator with anomalous dimension $\delta$ in the 4D CFT. In the RS picture it is generated by the GW scalar bulk mass. 

The novel pieces of the potential necessary for generating the Higgs-dependent second minimum arise from the dependence of the Higgs potential on the location of the IR brane $R'$. We assume that the Higgs field is sourced on the UV brane where the usual $\chi$-independent part of the potential in Eq.~\eqref{eq:Hpotential} arises from. The additional terms
\begin{equation}
 V_{H\chi}(\chi, H) = \lambda_2 |H|^2 \frac{\chi^{2+\alpha}}{k^{\alpha}} - \lambda_{H\epsilon} |H|^2 \frac{\chi^{2+\alpha+\epsilon}}{k^{\alpha+\epsilon}} - \lambda_4 |H|^4 \frac{\chi^{2\alpha}}{k^{2\alpha}} \label{eq:IRlocalized}
\end{equation}
arise from IR-localized interactions. Assuming the bulk mass of the Higgs is $m_b^2$ (in units of $R$) the Higgs VEV will scale as $z^{2\pm\sqrt{4+m_b^2}}$. We can easily show that the effect of a UV source $H_{\rm UV}$ will be a Higgs field that scales on the IR brane as $H_{\rm UV} \chi^{\sqrt{4+m_b^2}-2}=H_{\rm UV} \chi^{\frac{\alpha}{2}-1}$, where $\alpha = 2\sqrt{4+m_b^2}-2$. Hence, a brane-localized quadratic term will yield a localized potential $|H|^2 \chi^{2+\alpha}$, while a localized quartic will result in $|H|^4 \chi^{2\alpha}$. An additional dependence on the dilaton can be generated if we also introduce localized terms that include the GW scalar $\Phi \sim z^\epsilon$, or any other field with an approximately marginal dimension. This gives a modified quadratic term of the form $|H|^2 \chi^{2+\alpha+\epsilon}$, completing the terms outlined in Eq.~\eqref{eq:IRlocalized}.  

The detailed CFT interpretation of this mechanism is a spontaneously broken conformal sector, stabilized by the VEV of a marginal operator ${\cal O}_{GW}$, as in the standard GW stabilization of the dilaton. The ``techni-quarks'' of the CFT sector are charged under the EW gauge group and can form an $SU(2)$ doublet operator ${\cal O}_H$ of dimension $3+\alpha/2$ which couples linearly to a fundamental Higgs, i.e. $ {\cal O}_H^\dagger H $. If we also assume the presence of a marginal operator needed for GW stabilization ${\cal O}_{\epsilon}$ of dimension $4+\epsilon$, which may or may not be the same operator as ${\cal O}_{GW}$, we can have the following deformations in the UV action:
\begin{equation}
\tilde{\lambda}_H{\cal O}_H^\dagger H+ \tilde{\lambda}_\epsilon {\cal O}_{\epsilon}\, .
\end{equation}
The effective potential for these terms in the IR can be expanded as:
\begin{eqnarray}
V_{eff} &=& a_0 \chi^4 + a_1 \tilde{\lambda}^2_H H^2  \chi^{2+\alpha}  +a_2  \tilde{\lambda}^4_H H^4  \chi^{2\alpha} \nonumber\\ &+&a_3  \tilde{\lambda}_\epsilon \chi^{4+\epsilon}+ a_4  \tilde{\lambda}_\epsilon \tilde{\lambda}^2_H H^2   \chi^{2+\alpha+\epsilon} +...
\end{eqnarray}
where $a_i$ are the coefficients of the expansion. We see that we reproduce the terms in the potential of Eq.~\eqref{eq:IRlocalized}. In essence, the fundamental Higgs acts as an additional stabilizing force on the CFT, generating a second stable minimum for the dilaton.

Lastly, we assume that the couplings $\lambda$, $\lambda_{\rm GW}$, $\lambda_2$, $\lambda_{H\epsilon}$, and $\lambda_4$ are all positive.

 %%%%%%%%%%%%%%%%%%%%%%%%%%%%%%%%%%%%%%%%%%%%%%%%%%%%%%%%%%%%%%%%%%%%%%%%%%%
%%%%%%%%%%%%%%%%%%%%%%%%%%%%%%%%%%%%%%%%%%%%%%%%%%%%%%%%%%%%%%%%%%%%%%%%%%%
\section{Dynamics of the Dilaton-Higgs potential}
%%%%%%%%%%%%%%%%%%%%%%%%%%%%%%%%%%%%%%%%%%%%%%%%%%%%%%%%%%%%%%%%%%%%%%%%%%%
%%%%%%%%%%%%%%%%%%%%%%%%%%%%%%%%%%%%%%%%%%%%%%%%%%%%%%%%%%%%%%%%%%%%%%%%%%%

Let us now investigate the dynamics resulting from the Higgs-dilaton potential 
\begin{equation}
   V(\chi, H) = V_{\rm GW}(\chi) + V_{H\chi}(\chi, H) + V_H(H),
\end{equation}
where $V_{\rm GW}$ and $V_{H\chi}$ are given in Eqs.~\eqref{eq:GW}--\eqref{eq:IRlocalized} and $V_H$ is the UV brane-localized SM Higgs potential in Eq.~\eqref{eq:Hpotential}. To ensure that the Higgs mass is still dominated by $V_H$ we take the exponent  $\alpha$ in Eq.~\eqref{eq:IRlocalized} to be positive and not too large, implying $m_b^2 \simeq -3$ (and hence a Higgs field linear in $z$). 
We also take $V_{\rm GW}$ to be subdominant to $V_{H\chi}$ at small values of $\chi$, around the $V_{H\chi}$ minimum. 

\begin{figure}
\includegraphics[width=0.49\textwidth]{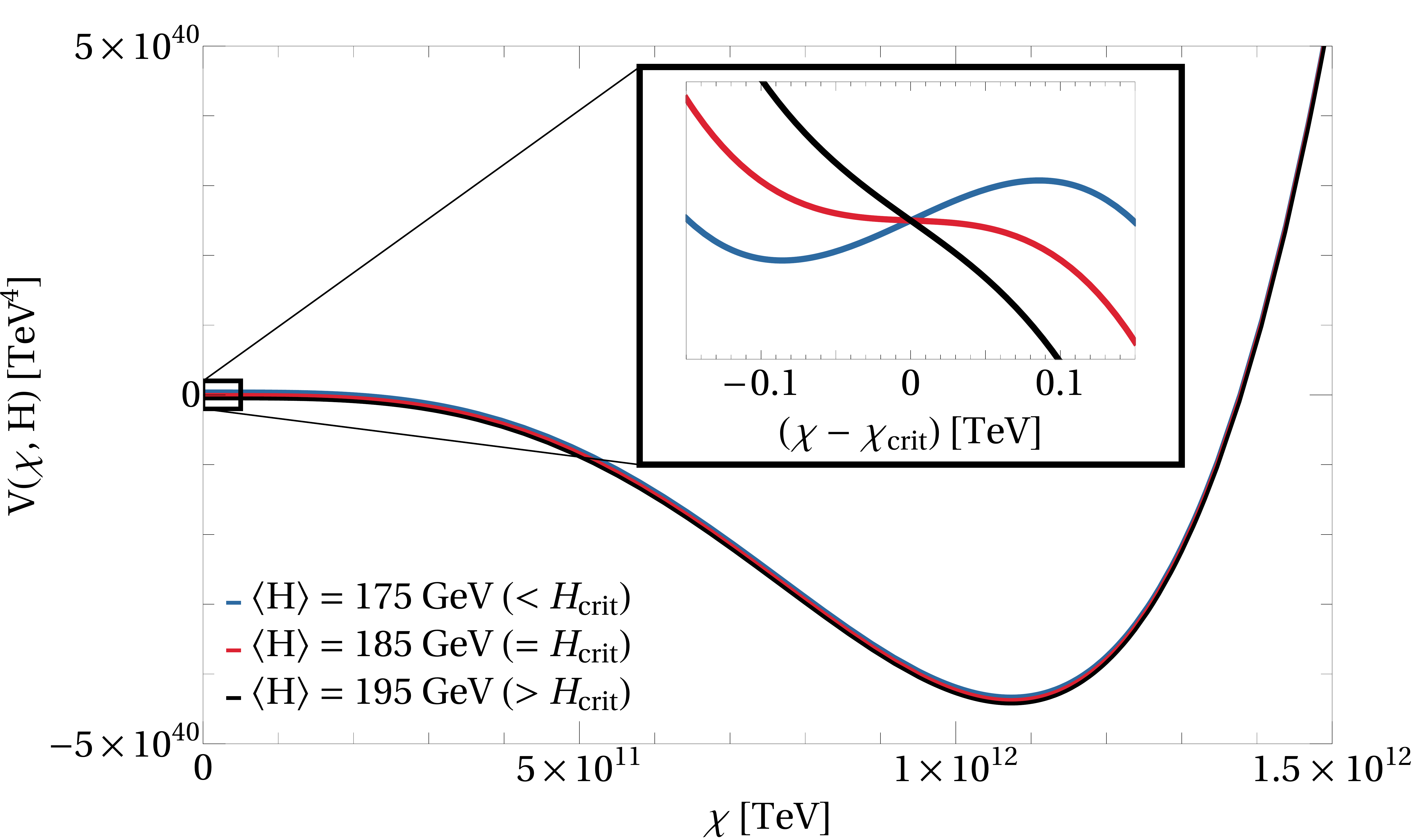}
    \caption{The dilaton potential $V(\chi, H) = V_{\rm GW}(\chi) + V_{H\chi}(\chi,H)$ for three different values of the Higgs VEV $\langle H \rangle$, using $k=10^{8}\TeV$, $\lambda_{\rm GW} = 1.2\times 10^{-5}$, $\lambda = 1.1\lambda_{\rm GW}$, $\lambda_2 = 0.005$, $\lambda_{H,\epsilon} = 0.018$, $\lambda_4 = 3$, $\delta = 0.01$, $\alpha = 0.05$, and $\epsilon = 0.1$. The true vacuum is depicted in the main figure while the second minimum is visible in the inset; note the potentials are shifted so that the inflection point $\chi_{\rm crit}$ lies at the origin. As the Higgs VEV is increased beyond $h_{\rm crit} = 185\GeV$ the second minimum disappears.}
    \label{fig:potential}
\end{figure}

For a finite range of Higgs values the $\chi$ potential admits two minima, one generated by $V_{\rm GW}$ and the other by $V_{H\chi}$. Above the critical value of the Higgs VEV $h_{\rm crit}$ the latter disappears, leaving only the GW minimum (see Fig.~\ref{fig:potential}). The minimum also disappears when the Higgs VEV is zero. However, to have a realistic cosmological history we need to modify the low $\chi$ behaviour of the potential. This generates a non-zero minimal value of the Higgs VEV $h_{\rm min}$ for the second potential minimum to exist. This modification does not affect any of the dynamics that we discuss below, and so we defer its study to the Cosmological Constraints Section.

The Higgs VEV in our part of the universe must be smaller than $h_{\rm crit}$, or the dilaton would have rolled down to the GW minimum, resulting in a crunch. We will show that there is a range of Higgs VEVs for which the metastable vacuum exists and survives until today. This range is close to $h_{\rm crit}$, hence the value of the Higgs VEV in our Hubble patch should lie just below $h_{\rm crit}$ without any associated tuning.

%%%%%%%%%%%%%%%%%%%%%%%%%%%%%%%%%%%%%%%%%%%%%%%%%%%%%%%%%%%%%%%%%
%%%%%%%%%%%%%%%%%%%%%%%%%%%%%%%%%%%%%%%%%%%%%%%%%%%%%%%%%%%%%%%%%
%%%%%%%%%%%%%%%%%%%%%%%%%%%%%%%%%%%%%%%%%%%%%%%%%%%%%%%%%%%%%%%%%
%%%%%%%%%%%%%%%%%%%%%%%%%%%%%%%%%%%%%%%%%%%%%%%%%%%%%%%%%%%%%%%%%

If we neglect $V_{\rm GW}$ at small $\chi$, $h_{\rm crit}$ can be computed by finding the value of $h$ for which $\partial_\chi V_{H\chi}$ has only one zero:
\begin{equation}\label{eq:hcrit}
    h_{\rm crit} = k \left(\frac{\lambda_2}{\lambda_{H\epsilon}}\frac{4-\alpha^2}{(2+\epsilon)^2-\alpha^2}\right)^{\frac{1-\alpha/2}{\epsilon}}\sqrt{\frac{\lambda_2}{\lambda_4}\frac{\epsilon (2+\alpha)}{2\alpha (2-\alpha+\epsilon)}}\ .
    \end{equation}
When $h=h_{\rm crit}$, $V_{\chi H}(\chi, h)$ has a single critical point (an inflection point), as shown in Fig.~\ref{fig:potential}, which lies at
\begin{equation}\label{eq:crit}
        \chi_{\rm crit}= k \left(\frac{\lambda_2}{\lambda_{H\epsilon}}\frac{4-\alpha^2}{(2+\epsilon)^2-\alpha^2}\right)^{1/\epsilon} \, .
\end{equation}
For $h\lesssim h_{\rm crit}$, we can also estimate the second minimum of the $\chi$ potential as 
\begin{equation}\label{eq:minimum}
        \chi_{\rm min}\simeq  \left(\frac{h^2}{k^\alpha}\frac{2\alpha \lambda_4}{(2+\alpha)\lambda_2}\right)^{\frac{1}{2-\alpha}} \end{equation}
 neglecting the $\lambda_{H\epsilon}$ term, which is suppressed at the minimum by $(\chi_{\rm min}/k)^\epsilon$ relative to $\lambda_2$.
 
 For small $\epsilon$ a mild hierarchy between couplings $\lambda_2 \lesssim \lambda_{H\epsilon}$ can generate a large hierarchy of scales $h_{\rm crit}, \chi_{\rm crit}, \chi_{\rm min} \ll k $. $\chi_{\rm min}$ sets the size of the extra dimension, hence determining the mass scale of new states potentially observable at colliders. The little hierarchy problem is reflected in the need to impose a mild hierarchy between $h$ and $\chi_{\rm min}$
\begin{equation}
    \frac{h}{\chi_{\rm min}}\simeq \frac{h_{\rm crit}}{\chi_{\rm min}}\lesssim 0.1\, .
    \end{equation}
This implies a hierarchy of couplings $\lambda_2, \lambda_{H\epsilon}< 10^{-2} \alpha \lambda_4$ that we discuss in detail in the Section dedicated to NDA.

The most interesting consequence of this little hierarchy is the prediction of a light dilaton. Its mass for small $\epsilon$, $\alpha$ and $\delta$ is
\begin{equation}
m_\chi \simeq m_h \sqrt{\frac{h}{\chi_{\rm min}}\frac{\pi \sin\theta}{\sqrt{6}N}-\frac{8\pi^2(\lambda-\lambda_{\rm GW})}{N^2}\frac{\chi_{\rm min}^2}{m_h^2}}\, ,
    \label{eq:dilaton_mass_precise}
\end{equation}
where 
\be
\sin\theta \sim \frac{(\lambda_2-\lambda_{H\epsilon})}{N}\frac{h \chi_{\rm min}}{m_h^2}
\ee
parametrizes the dilaton mixing with the Higgs and we have included the contribution of $V_{\rm GW}$ to the dilaton mass. We explore the dilaton phenomenology in the next Section.

As stated above, the previous analysis is valid only if the GW potential is subleading to $V_{\chi H}$ around $\chi_{\rm crit}$, leading to an upper bound 
\begin{equation}\label{eq:GWconstraint}
    \lambda\sim \lambda_{\rm GW} \lesssim  \frac{\lambda_2^2}{\lambda_4} .
\end{equation}
If $V_{\rm GW}$ dominates over $V_{\chi H}$ at $\chi_{\rm crit}$, it washes out the metastable minimum.

%%%%%%%%%%%%%%%%%%%%%%%%%%%%%%%%%%%%%%%%%%%%%%%%%%%%%%%%%%%%%%%%%%%%%%%%%%%
%%%%%%%%%%%%%%%%%%%%%%%%%%%%%%%%%%%%%%%%%%%%%%%%%%%%%%%%%%%%%%%%%%%%%%%%%%%
\section{Phenomenological Consequences}
 %%%%%%%%%%%%%%%%%%%%%%%%%%%%%%%%%%%%%%%%%%%%%%%%%%%%%%%%%%%%%%%%%%%%%%%%%%%
%%%%%%%%%%%%%%%%%%%%%%%%%%%%%%%%%%%%%%%%%%%%%%%%%%%%%%%%%%%%%%%%%%%%%%%%%%%

While we are dealing with a warped 5D model, the essence of our mechanism for a light Higgs (and hence the experimental predictions) is completely different from a vanilla holographic composite Higgs model. Our theory does not have top partners, light or heavy; they play no role in the stabilization of the Higgs hierarchy. There are no KK gluons either. There have to be KK electroweak gauge bosons, since the Higgs propagates in the bulk, but they do not have to be light and also play no role in stabilizing the hierarchy. The Higgs gets a large fraction of its potential on the UV brane, and can be thought of as a mixture of elementary and composite states. 

The most salient phenomenological feature of our model is the existence of a light dilaton, as shown in Eq.~\eqref{eq:dilaton_mass_precise}. 
Due to its mixing with the Higgs it inherits all SM Higgs couplings suppressed by the mixing angle $\theta$. In addition to these, the dilaton has direct couplings to the SM fields. Since the SM fermions are assumed to be localized on the UV brane and the dilaton is localized predominantly on the IR brane their direct couplings are negligible. In contrast, electroweak gauge bosons propagate in the bulk and their direct coupling to the dilaton is given by~\cite{Csaki:2000zn,Csaki:2007ns}
\begin{equation}
\frac{\chi}{2 \chi_{\rm min} \log \frac{R'}{R} } (F_{\mu\nu}^2+Z_{\mu\nu}^2 + 2 W_{\mu\nu}^2) \ .
\label{eq:directcoupling}
\end{equation}
The direct couplings to the $Z,W$ mass terms are a small correction to those obtained from the mixing with the Higgs, and their effects can be neglected.

The mass of the dilaton has a lower bound determined by the contribution from $V_{\rm GW}$. Given that we need $\lambda_{\rm GW}\lesssim \lambda$ to have a second minimum at large values of $\chi$, the $V_{\rm GW}$ contribution to the dilaton mass is always negative at the metastable minimum. Therefore, if we do not tune the two terms in Eq.~\eqref{eq:dilaton_mass_precise}, $m_\chi>0$ implies
\be
m_\chi \gtrsim 2\pi \frac{\chi_{\rm min}}{N}\sqrt{2(\lambda-\lambda_{\rm GW})}\, .
\ee
Numerically $\chi_{\rm min}\simeq {\rm TeV}$, $N\lesssim 40$ and $\lambda, \lambda_{\rm GW} \gtrsim 10^{-6}$, from the arguments in the next Section, so we expect a lower bound of $\mathcal{O}(100)$~MeV. We also have an upper bound that can be easily obtained from Eq.~\eqref{eq:dilaton_mass_precise}: $m_\chi \lesssim 0.2 m_h$. This bound is harder to saturate because the maximum value of $\sin \theta$ is limited by the need to take $\lambda_2$ to be $\mathcal{O}(10^{-2})$. As stated earlier, this requirement arises from the little hierarchy problem, which we explore in detail in the next Section.

In summary, we have a dilaton with mass $0.1\;{\rm GeV}\lesssim m_\chi \lesssim 10$~GeV and couplings to fermions proportional to $\sin\theta \sim m_\chi^2/m_h^2$. The direct coupling to photons in Eq.~\eqref{eq:directcoupling} plays an important role in its phenomenology, giving an $\mathcal{O}(1)$ correction to its branching ratios.

To explore the properties of this dilaton and the experimental constraints on it, we randomly generated $10^5$ points in the parameter space, fixing the parameters $k = 10^{11}$~GeV, $\delta = 0.01$, $N=3$ and $\alpha = 0.05$, while uniformly sampling the other parameters from the ranges $\lambda_{\rm GW} \in (0.5, 1.5) \times 10^{-5}$, $\lambda_2 \in (0.5, 1.5) \times 10^{-2}$, $\lambda_{H\epsilon} \in (2, 4) \cdot \lambda_2$, $\lambda_4 \in (2, 3)$, and $\epsilon \in (0.03, 0.1)$. We also took $\lambda = 1.1\, \lambda_{\rm GW}$ and set the Higgs VEV $\langle H \rangle \simeq 174 \GeV$. The parameter values chosen here reflect the little hierarchy $h_{\rm crit}/\chi_{\rm min}\lesssim 0.1$ and $V_{\rm GW}\lesssim V_{H\chi}$ at $\chi_{\rm min}$. This is explained further in the next Section. To probe lower dilaton masses, we performed a similar analysis of $5 \times 10^4$ points, choosing $N=8$, $\alpha = 0.1$, $\lambda_{\rm GW} = 2\times 10^{-6}$, $\lambda_2 \in (0.5, 1) \times 10^{-2}$, and $\epsilon \in (0.05, 0.1)$, while keeping the other parameters the same. 
Points were excluded from our analysis if they failed to satisfy the following four criteria:
\begin{itemize}
\item The metastable vacuum must exist and be located at $\chi_{\rm crit} > 1$~TeV.
\item $h_{\rm crit} \leq 2\TeV$ so the Higgs VEV is natural.
\item The metastable vacuum reproduces the SM values of the Higgs mass and VEV and corresponds to a stable local minimum of the 2 dimensional potential. 
\item The O(4) bounce action $S_4$ between the two potential minima is at least $\mathcal{O}(200)$ so that tunnelling is suppressed.
\end{itemize}

The bounce action (see~\cite{Coleman:1977py}) was computed by numerically solving the Euclidean equation of motion, using the shooting method to satisfy the boundary conditions. Due to the normalization of the dilaton kinetic term, $S_4$ scales with $N$ as $(N^2 - 1)^2$. In practice the bounce action is quite large: even for our smallest choice of $N=3$ (set by perturbativity of 5D gravity, discussed in the next Section), it is at least $\mathcal{O}(10^4)$ for points that satisfy the other three criteria. Tunnelling to the GW minimum is therefore sufficiently suppressed for these points, with any value of $N$.

\begin{figure*}
    \includegraphics[width=\textwidth]{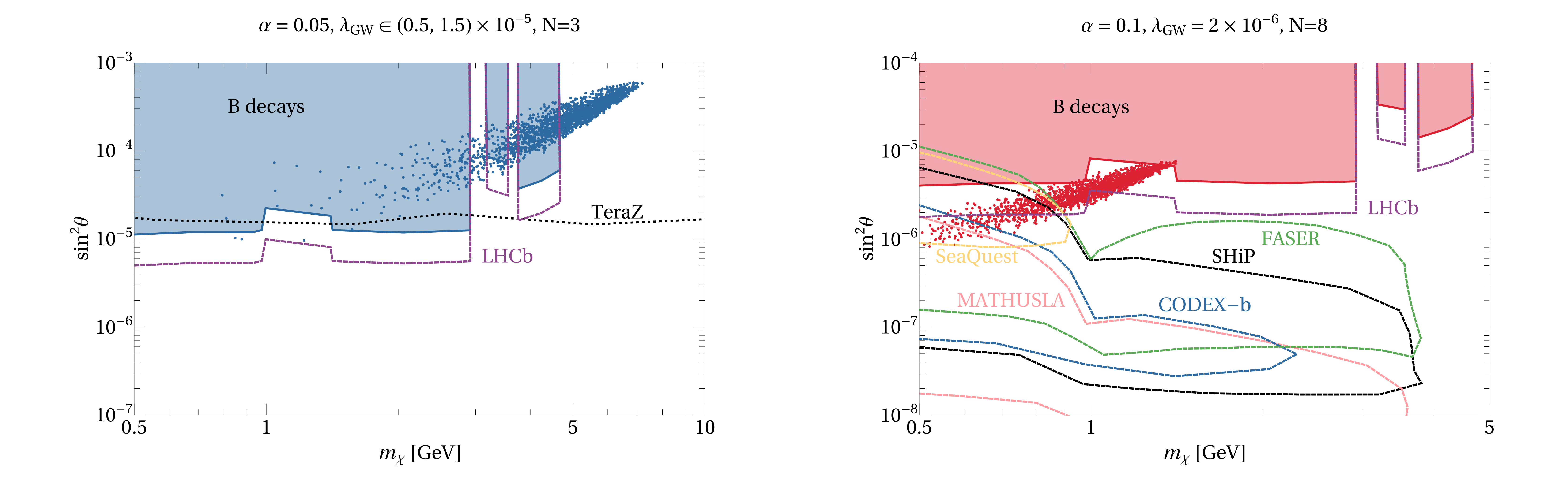}
    \caption{The dilaton mass $m_\chi$ and its mixing angle with the Higgs $\sin^2 \theta$ for randomly sampled points from our model. We use different choices of parameter ranges in the two panels, which are detailed in the main body of the text. We show current bounds from $B$ meson decays at LHCb~\cite{Aaij:2012vr,Aaij:2015tna} (blue and red shaded regions), adapted from \cite{Flacke:2016szy,Frugiuele:2018coc}. We estimate updated $B$ decay bounds following Run 3 of the LHC (purple lines), assuming an integrated luminosity of $15{\rm~fb}^{-1}$~\cite{Beacham:2019nyx}. We also include projections for bounds from SHiP~\cite{Alekhin:2015byh}, MATHUSLA~\cite{Chou:2016lxi,Evans:2017lvd}, CODEX-b~\cite{Gligorov:2017nwh,Evans:2017kti}, FASER~\cite{Feng:2017uoz,Feng:2017vli}, and SeaQuest~\cite{Gardner:2015wea,Beacham:2019nyx} (dashed lines, right), as well as from $Z \to Z^* \chi$ and $e^+ e^- \to Z \chi$ at the FCCee running on the $Z$ pole (Tera-$Z$)~\cite{Frugiuele:2018coc} (dotted black line, left), rescaled from the corresponding LEP limits~\cite{Acciarri:1996um,Schael:2006cr}.}
    \label{fig:massmix}
\end{figure*}

The results of the two scans are plotted in Fig.~\ref{fig:massmix}. We indicate the relevant experimental bounds from rare $B$ meson decays~\cite{Aaij:2012vr,Aaij:2015tna}, adapted from \cite{Flacke:2016szy,Frugiuele:2018coc}. The strength of this constraint depends on the branching ratio of the dilaton to leptons, which is suppressed by the direct coupling to photons. Once the kinetic term normalization is taken into account, this coupling scales as $1/\sqrt{N^2-1}$ (see Eq.~\eqref{eq:directcoupling}), and hence the $B$ decay bound is stronger for $N=8$ than $N=3$. The region in the $m_\chi$--$\,\sin\theta$ plane populated by the two scans can be understood from Eq.~\eqref{eq:dilaton_mass_precise}: the points approximately fall on the curve $m_\chi \sim \sqrt{\sin\theta}$, with upper (lower) bounds determined by the values of $\lambda_{2, H\epsilon}$ ($\lambda,\lambda_{\rm GW}$).

There are two regions in the parameter space free of bounds around $0.5$--$1.5$~GeV and $5$--$7$~GeV. These could be probed at LHCb with more $B$ decay data, which will be collected during Run 3 of the LHC; at future searches for hidden, light particles, particularly the beam dump experiments SHiP~\cite{Alekhin:2015byh} and SeaQuest~\cite{Gardner:2015wea,Beacham:2019nyx} and the collider experiments FASER~\cite{Feng:2017uoz,Feng:2017vli}, CODEX-b~\cite{Gligorov:2017nwh,Evans:2017kti} and MATHUSLA~\cite{Chou:2016lxi,Evans:2017lvd}; and with a Tera-$Z$ program at future lepton colliders such as the FCCee or CepC~\cite{Frugiuele:2018coc,Acciarri:1996um,Schael:2006cr,Abada:2019zxq}, as illustrated in Fig.~\ref{fig:massmix}.

\begin{figure}
    \includegraphics[width=0.49\textwidth]{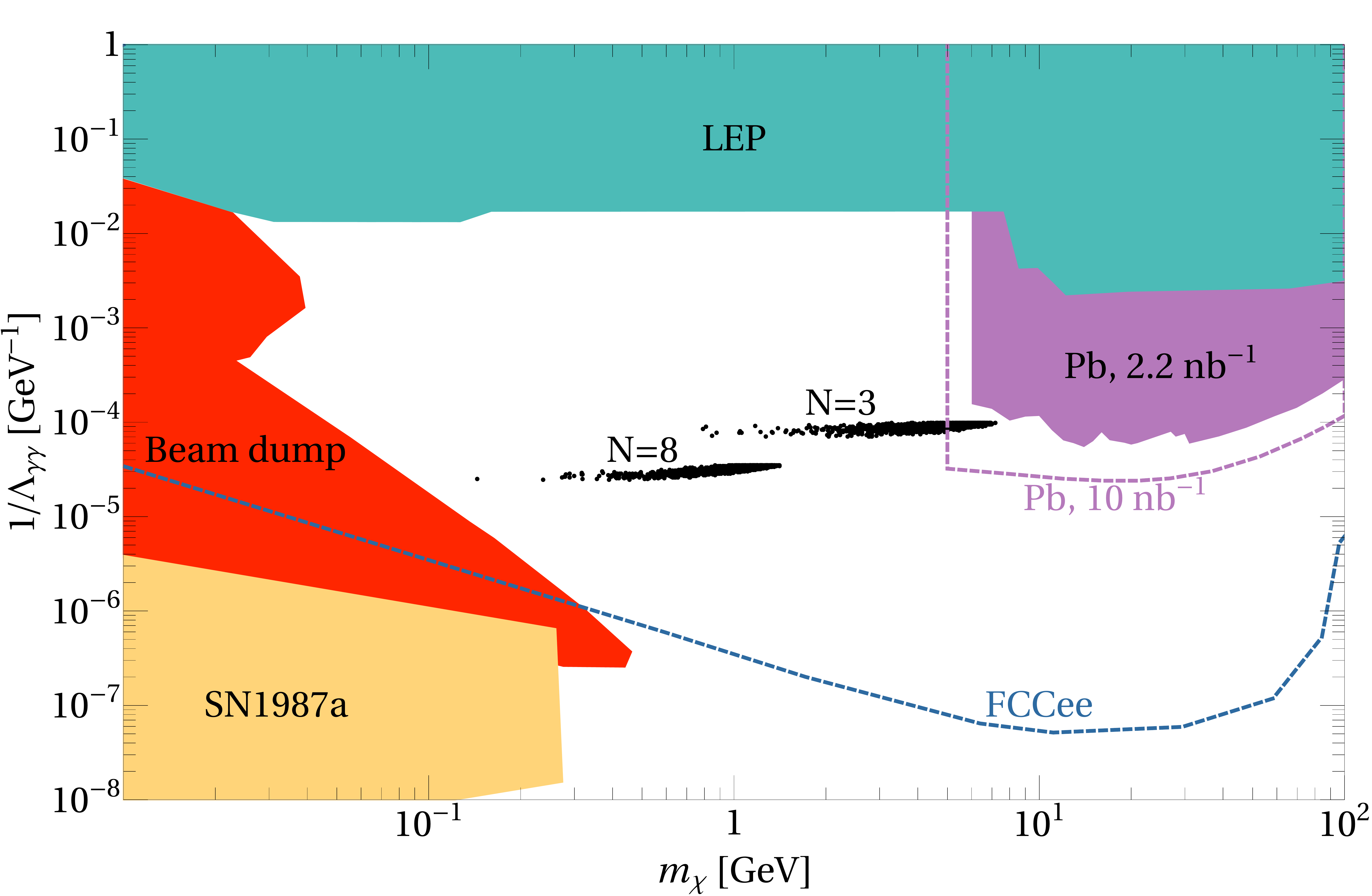}
    \caption{The dilaton mass $m_\chi$ and its coupling to photons $1/\Lambda_{\gamma\gamma}$ for randomly sampled points from our model. We include points from the two choices of parameter ranges in Fig.~\ref{fig:massmix}. We show bounds, adapted from \cite{Bauer:2018uxu}, from searches at LEP for $e^+e^- \to \gamma \chi \to 3\gamma$~\cite{Mimasu:2014nea,Jaeckel:2015jla} (turquoise), beam dump experiments~\cite{Bjorken:1988as} (red), supernova SN1987a~\cite{Payez:2014xsa,Jaeckel:2017tud} (orange), and $\gamma\gamma \to \chi \to \gamma \gamma$ in lead ion collisions at the LHC with $2.2{\rm~nb}^{{\rm-1}}$ of data~\cite{ATLAS:2020khf} (purple). We also include projections for an FCCee search for $e^+e^- \to \gamma \chi \to 3\gamma$~\cite{Bauer:2018uxu,Bauer:2017ris} (blue line) and for lead ion collisions with $10{\rm~nb}^{{\rm-1}}$ of data~\cite{Knapen:2016moh,Knapen:2017ebd} (purple line).}
    \label{fig:photoncoups}
\end{figure}

Finally, in Fig.~\ref{fig:photoncoups} we plot the dilaton coupling to photons $1/\Lambda_{\gamma\gamma}$ for each point in our scans, alongside current and projected experimental bounds adapted from \cite{Bauer:2018uxu}. These bounds provide constraints on the model that are independent of $\sin\theta$. We normalize the coupling such that the dilaton-photon interaction term is $\frac{1}{4\Lambda_{\gamma\gamma}} \tilde \chi F_{\mu\nu}^2$, with $\tilde \chi$ the canonically normalized dilaton. The region of parameter space populated by our model evades the existing bound from LEP searches for $e^+e^- \to \gamma \chi \to 3 \gamma$~\cite{Mimasu:2014nea,Jaeckel:2015jla}, but it could be probed at a future lepton collider through the same search channel~\cite{Bauer:2018uxu,Bauer:2017ris}.
Note that the latter bound (Fig.~\ref{fig:photoncoups}, blue line) assumes a 100\% branching ratio of the dilaton to photons. This is not exactly true, but the branching ratio is always at least $\sim 10\%$ and the strength of the bound scales as its square root; thus, even after taking this effect into account, a future lepton collider would indeed probe our model.
Lead ion collisions at the LHC also constrain the parameter space for $m_\chi \gtrsim 5$~GeV, through searches for $\gamma\gamma \to \chi \to \gamma \gamma$ in peripheral collisons. At higher luminosity ($10{\rm~nb}^{{\rm-1}}$) this bound would be sensitive to our model~\cite{Knapen:2016moh,Knapen:2017ebd}, while the current bound~\cite{ATLAS:2020khf} does not reach the sensitivity needed. 

%%%%%%%%%%%%%%%%%%%%%%%%%%%%%%%%%%%%%%%%%%%%%%%%%%%%%%%%%%%%%%%%%
%%%%%%%%%%%%%%%%%%%%%%%%%%%%%%%%%%%%%%%%%%%%%%%%%%%%%%%%%%%%%%%%%
%%%%%%%%%%%%%%%%%%%%%%%%%%%%%%%%%%%%%%%%%%%%%%%%%%%%%%%%%%%%%%%%%
%%%%%%%%%%%%%%%%%%%%%%%%%%%%%%%%%%%%%%%%%%%%%%%%%%%%%%%%%%%%%%%%%

%%%%%%%%%%%%%%%%%%%%%%%%%%%%%%%%%%%%%%%%%%%%%%%%%%%%%%%%%%%%%%%%%
%%%%%%%%%%%%%%%%%%%%%%%%%%%%%%%%%%%%%%%%%%%%%%%%%%%%%%%%%%%%%%%%%
%%%%%%%%%%%%%%%%%%%%%%%%%%%%%%%%%%%%%%%%%%%%%%%%%%%%%%%%%%%%%%%%%
%%%%%%%%%%%%%%%%%%%%%%%%%%%%%%%%%%%%%%%%%%%%%%%%%%%%%%%%%%%%%%%%%

%%%%%%%%%%%%%%%%%%%%%%%%%%%%%%%%%%%%%%%%%%%%%%%%%%%%%%%%%%%%%%%%%%%%%%%%%%%
%%%%%%%%%%%%%%%%%%%%%%%%%%%%%%%%%%%%%%%%%%%%%%%%%%%%%%%%%%%%%%%%%%%%%%%%%%%
\label{sec:NDA}
\section{NDA and tuning}
%%%%%%%%%%%%%%%%%%%%%%%%%%%%%%%%%%%%%%%%%%%%%%%%%%%%%%%%%%%%%%%%%%%%%%%%%%%
%%%%%%%%%%%%%%%%%%%%%%%%%%%%%%%%%%%%%%%%%%%%%%%%%%%%%%%%%%%%%%%%%%%%%%%%%%%

We have seen that the phenomenologically successful models require small values of the couplings $\lambda_2, \lambda$ and $\lambda_{\rm GW}$. The reason behind this is simple: there is an upper limit on the brane-localized Higgs quartic $\lambda_4 \lsim 3$ imposed by requiring that a Landau pole does not appear before we hit at least a few KK modes. The value of $\chi$ at the metastable minimum must be larger than 1 TeV to avoid LHC bounds~\cite{Aaboud:2018bun,Sirunyan:2019vgt}, leading to $\lambda_4/\lambda_2 \gsim 10^2$ (see Eq.~\eqref{eq:minimum}), and hence $\lambda_2 \lsim 10^{-2}$. Finally, we must ensure that the GW part of the potential does not overwhelm the second minimum. In accordance with Eq.~\eqref{eq:GWconstraint} this yields $\lambda ,\lambda_{\rm GW} \lsim 10^{-5}$. These values are quite a bit smaller than one would expect from simple NDA in a warped extra dimension. For example, $\lambda_2$ arises from a brane-localized mass term for the bulk Higgs scalar, which is expected to be quadratically divergent, leading to $\lambda_2 \sim \frac{1}{16\pi^2} \frac{\Lambda^2}{\chi^2}$, where $\Lambda$ is the local cutoff.\footnote{Note that $\lambda_2$ will also get a contribution from bulk loops since it corresponds to a mismatch between bulk and brane terms. However, since the brane term dominates the NDA, we restrict ourselves to that.} Similarly, the GW quartic $\lambda$ has been estimated in~\cite{Bellazzini:2012vz} and is given by $\lambda \sim \frac{1}{16\pi^2} \frac{\Lambda^4}{\chi^4}$. We can see that in order to minimize the tuning we should lower the local cutoff scale such that $\Lambda \lsim \chi$. In this case the natural value for $\lambda_2$ will be approximately $\sim 10^{-2}$ as needed. Depending on the actual value of $\Lambda \lsim \chi$ there may be some tuning still left in the  the GW potential. For example, if $\Lambda \sim \chi$ we get a tuning of around a percent corresponding to the usual little hierarchy problem. If the Higgs VEV was a factor of $\sim 10$ larger the NDA value for $\lambda$ would have been sufficently small since $\chi$ could also have increased. 

Whether one can achieve  $\Lambda /\chi \lsim 1$ depends on the particular UV completion of the theory. One simple way to do it is to impose that the bulk is supersymmetric, while SUSY is broken on the UV brane (and also spontaneously broken on the IR brane to allow the generation of the dilaton quartic $\chi^4$). This would be along the lines of~\cite{Sundrum:2009gv}, and imply that the fields in the bulk have light superpartners. While a complete discussion of such a setup is beyond the scope of this paper, we outline a simple scenario without tuning. A high SUSY breaking scale on the UV brane ensures that all sfermions are ultra-heavy. The structure of SUSY breaking on the UV brane keeps the electroweak gauginos and Higgsinos light, of order a few hundred GeV, as in split SUSY models.  In this case the IR brane cutoff relevant for $\lambda$ would be lowered to the gauino mass scale and $\lambda \simeq 10^{-5}$ would be fully natural. 

For completeness let us also discuss the bounds on $N$, which are important for establishing the phenomenologically viable regions of the dilaton mass and mixing angle. Requiring that the 5D gravitational theory is not strongly coupled (i.e. the 5D AdS scalar curvature $20/R^2$ is smaller than the 5D cutoff $\Lambda^3 \sim 24\pi^3 M_*^3$) yields $N\gsim 3$. Assuming that the gravitational KK modes have approximately the same $1/N$ suppression as the weak gauge KK modes, and using the matching of weak gauge couplings, we get an upper bound $N\lsim 40$ .
 
%%%%%%%%%%%%%%%%%%%%%%%%%%%%%%%%%%%%%%%%%%%%%%%%%%%%%%%%%%%%%%%%%%%%%%%%%%%
%%%%%%%%%%%%%%%%%%%%%%%%%%%%%%%%%%%%%%%%%%%%%%%%%%%%%%%%%%%%%%%%%%%%%%%%%%%
\section{Cosmological Constraints }
%%%%%%%%%%%%%%%%%%%%%%%%%%%%%%%%%%%%%%%%%%%%%%%%%%%%%%%%%%%%%%%%%%%%%%%%%%%
%%%%%%%%%%%%%%%%%%%%%%%%%%%%%%%%%%%%%%%%%%%%%%%%%%%%%%%%%%%%%%%%%%%%%%%%%%%

Finally, we consider the cosmological constraints on our model. First, as outlined in Eq.~\eqref{eq:hrange}, we need to ensure that the Hubble scale during inflation is below the electroweak scale, so that the dilaton potential can be sensitive to Higgs VEVs of ${\cal O}({\rm TeV})$. This requires the scale of inflation $M_I$ to be below the intermediate scale $\sqrt{M_W M_{\rm Pl}} \simeq 10^7$ TeV. Generically, to avoid a contribution to the vacuum energy from the UV completion of the Higgs sector which dominates over the inflaton sector, we require $\Lambda < M_I < 10^7$ TeV, where $\Lambda$ is the cutoff of the Higgs sector. (However, recall from the previous Section that to have a natural Higgs-dilaton potential new states carrying EW charges must appear well below this scale.)
We also require that the energy density in the true vacuum is indeed always negative---i.e. $\lambda \chi^4_{GW} > M_I^4$, which results in $k\gtrsim 17 M_I$ for $\lambda \simeq 10^{-5}$. In addition,  if we assume that the cosmological constant problem is solved via a standard anthropic mechanism, we need to require that the highest possible CC in the landscape is below $\sim \lambda \chi^4_{GW}$, so that all patches reaching the true minimum of the GW potential crunch. This condition may naturally arise for an ${\cal O}(\Lambda)$ SUSY breaking scale on the UV brane.

A second important requirement is that quantum diffusion never dominates over classical evolution. This prevents patches with the wrong value of $h$ to enter a phase of eternal inflation. We want to ensure that for a Higgs VEV large enough or small enough such that the dilaton potential has only one minimum with a negative CC, the dilaton does indeed roll to that minimum during inflation and does not get stuck in an eternally inflating phase. For large Higgs VEVs, the second derivative of the dilaton potential is always at least of $\mathcal{O}(v^2\simeq (174\;{\rm GeV})^2$) (see Eq.~\eqref{eq:IRlocalized}) and so classical rolling dominates over quantum diffusion, since we already imposed that the Hubble scale is less than the electroweak scale. 

The case of a small or vanishing Higgs VEV could potentially be more problematic. Once the Higgs VEV is zero, the $\chi$ potential is a pure $\chi^4$ term even for very small values of $\chi$, leading to a very small second derivative near the origin. Hence for any choice of the Hubble scale during inflation there are regions which support eternal inflation. This can be avoided by a simple modification of the model following the ideas of~\cite{vonHarling:2017yew,Baratella:2018pxi}, which was also used in~\cite{Bloch:2019bvc}. We include an additional term $\lambda_{\gamma}\chi^{\gamma}\tilde{\Lambda}^{4-\gamma}$ in the $\chi$ potential, corresponding to explicit breaking of scale invariance at the scale $\tilde{\Lambda} \ll k$. 
This term does not change our analysis for  $\chi \gg \chi_*$, defined as the scale where the effective quartic coupling blows up, i.e. $ \chi_*   \sim \tilde{\Lambda} \lambda_\gamma^{\frac{1}{4-\gamma}}$. 

At $\chi \lesssim \chi_*$ the potential is dominated by the explicit breaking term, which signals that for $\chi \lesssim \chi_*$  the description in terms of a dilaton breaks down. In that region we expect the effective potential to be dominated by the mass scale $\chi_*$, and effectively behave as if a negative mass term of order $\chi^2_*$ was generated. Such an explicit breaking term can be generated by any relevant operator which has a negligible coupling in the UV and grows to be ${\cal O}(1)$ in the IR.  One such realization could be to have the SM QCD (or a BSM QCD-like gauge theory) in the bulk. To see how this term emerges we can find the RG evolution of the gauge coupling for the group in the bulk, assuming the presence of a UV and IR brane~\cite{vonHarling:2017yew,Csaki:2007ns}:
 \begin{equation}
\label{eq:QCDbetafunc}
\frac{1}{g^{2}(Q, \chi)}=\frac{\log \frac{k}{\chi}}{k g_{5}^{2}}-\frac{b_{\mathrm{UV}}}{8 \pi^{2}} \log \frac{k}{Q}-\frac{b_{\mathrm{IR}}}{8 \pi^{2}} \log \frac{\chi}{Q}+\tau
\end{equation}
where $Q$ is the running scale and the dependence on $\chi$ is introduced due to the finite size of the extra dimension. $\tau=\tau_{\rm UV}+\tau_{\rm IR}$ contains the brane-localized kinetic terms and $b_{UV,IR}$ are the 4D $\beta$-functions on the two branes. Note that Eq.~\eqref{eq:QCDbetafunc} is valid only for $Q<\chi$. From Eq.~\eqref{eq:QCDbetafunc} we get the $\chi$ dependence of the dynamical scale of the bulk gauge group 
\begin{equation}
\begin{split}
    \tilde \Lambda(\chi) &= \left(k^{b_{\mathrm{UV}}} \chi^{b_{\mathrm{IR}}} e^{-8 \pi^{2} \tau}\left(\frac{\chi}{k}\right)^{-b_{\mathrm{CFT}}}\right)^{\frac{1}{b_{\mathrm{UV}}+b_{\mathrm{IR}}}} \\
    &=\Lambda_{0}\left(\frac{\chi}{\chi_{\min }}\right)^{n} \, .
\end{split}
\end{equation}
For our benchmark point of $\chi_{\rm min}\simeq 1$ TeV and $\langle H\rangle =0$, QCD in the bulk gives $\tilde \Lambda(\chi_{\rm min})\sim \Lambda_{\rm QCD} \sim 100$ MeV. The lower bound on $n$ is $n\gsim0.1$, which results in $\chi_* \sim 10-100 \MeV \sim \Lambda_{\rm QCD}$.  To avoid eternal inflation we have to take the highest Hubble constant in the landscape to be just below $\chi_*$. Assuming that the corresponding cutoff also sets the scale of the CC landscape, we find that $\Lambda<\sqrt{\chi_* M_{\rm Pl}} \lsim 10^5$ TeV. If we use  a different relevant operator, e.g. by having a dark gauge group with a higher confining scale, the cutoff can be taken all the way to $10^7$ TeV. This addition also generates a minimal value for the Higgs VEV $h_{\rm min}$ below which the universe will crunch. This will occur when the corresponding $\chi_{\rm min}$ is of order $\chi_*$, and so $h_{\rm min} \sim 0.1\chi_*$ (see Eq.~\eqref{eq:minimum}). Note that including QCD in the bulk has also an impact on the light dilaton phenomenology whose study we leave to future work.

Thus we see that our mechanism can successfully avoid eternal inflation as long as QCD or another QCD-like group is in the bulk. The resulting upper bound on the cutoff is $10^5$--$10^7$ TeV, depending on the implementation, and our solution is compatible with the usual anthropic solution of the CC problem. This gauge dynamics in the bulk also ensures that even if the CFT goes to the high temperature symmetric phase after reheating, which is  the case for a high reheating temperature, it will transition to the Higgs generated minimum when the temperature is close to $\chi_*$. It would be interesting to study the cosmological implications of this phase transition not far from BBN (for the study of the dilatonic phase transition see e.g. \cite{vonHarling:2017yew,Baratella:2018pxi,Konstandin:2011dr,Agashe:2019lhy,Creminelli:2001th}).

%%%%%%%%%%%%%%%%%%%%%%%%%%%%%%%%%%%%%%%%%%%%%%%%%%%%%%%%%%%%%%%%%%%%%%%%%%%
%%%%%%%%%%%%%%%%%%%%%%%%%%%%%%%%%%%%%%%%%%%%%%%%%%%%%%%%%%%%%%%%%%%%%%%%%%%
\section{Summary and Outlook}
%%%%%%%%%%%%%%%%%%%%%%%%%%%%%%%%%%%%%%%%%%%%%%%%%%%%%%%%%%%%%%%%%%%%%%%%%%%
%%%%%%%%%%%%%%%%%%%%%%%%%%%%%%%%%%%%%%%%%%%%%%%%%%%%%%%%%%%%%%%%%%%%%%%%%%%

We have presented a new approach to the electroweak hierarchy problem, connecting the Higgs mass to cosmological stability and the possibility of viable inflationary and post-inflationary expansion. In this mechanism the Higgs and electroweak gauge fields live in the bulk of a 5D RS model. However, unlike the original RS construction~\cite{Randall:1999ee} and subsequent composite Higgs models~\cite{Bellazzini:2014yua, Panico:2015jxa},  a sizeable fraction of the Higgs mass arises from the UV brane and strongly affects the dynamics of the dilaton. In particular, a second, shallower minimum for the dilaton, with a non-negative CC, is generated in addition to the regular GW-stabilized true vacuum. 
In the true vacuum the CC is assumed to be large and negative, leading to a rapid crunch of any patch in this minimum.

A second minimum is only generated when the Higgs VEV is between the Hubble scale at inflation and the TeV scale. As a consequence, only patches with $m_H^2<0$ and $|m_H^2|\lesssim {\rm TeV}^2$ survive inflation and expand, giving rise to the Universe that we observe today. Observers at the weak scale initially conclude that the Higgs mass is unnaturally small, but by probing higher energies they can find that it is selected by the dynamics of the Higgs-dilaton potential during inflation.

The most distinctive prediction is a light dilaton with mass below 10 GeV and above 100 MeV,  a consequence of the little hierarchy between the EW scale and the TeV scale enforced by LHC constraints. Part of the dilaton parameter space is already probed by existing data from colliders, specifically measurements of rare $B$ decays~\cite{Aaij:2012vr,Aaij:2015tna}. The remaining parameter space can be probed by a future $e^+ e^-$ collider~\cite{Acciarri:1996um,Schael:2006cr,Abada:2019zxq} and proposed searches for light, weakly-coupled particles~\cite{Alekhin:2015byh,Chou:2016lxi,Evans:2017lvd,Gligorov:2017nwh,Evans:2017kti,Feng:2017uoz,Feng:2017vli,Gardner:2015wea,Beacham:2019nyx}. Additionally, we expect new gauge and Higgs KK states to appear at the TeV scale which may be accompanied by gauge and Higgs super-partners, which, however, play no role in canceling quadratic divergences. The physics of this UV completion is left for future studies, but it could lead to observable signals at the HL-LHC.

%%%%%%%%%%%%%%%%%%%%%%%%%%%%%%%%%%%%%%%%%%%%%%%%%%%%%%%%%%%%%%%%%%%%%%%%%%%
%%%%%%%%%%%%%%%%%%%%%%%%%%%%%%%%%%%%%%%%%%%%%%%%%%%%%%%%%%%%%%%%%%%%%%%%%%%
\mySections{Acknowledgments}
We thank Brando Bellazzini, Raman Sundrum, and Tomer Volansky for useful comments. 
 C.C. and A.I. are supported in part by the NSF grant PHY-1719877. C.C. is also supported in part by the BSF grant 2016153. M.G. is supported in part by the Israel Science Foundation (Grant No. 1302/19). 
\bibliography{references}{}

\begin{thebibliography}{59}%
\makeatletter
\providecommand \@ifxundefined [1]{%
 \@ifx{#1\undefined}
}%
\providecommand \@ifnum [1]{%
 \ifnum #1\expandafter \@firstoftwo
 \else \expandafter \@secondoftwo
 \fi
}%
\providecommand \@ifx [1]{%
 \ifx #1\expandafter \@firstoftwo
 \else \expandafter \@secondoftwo
 \fi
}%
\providecommand \natexlab [1]{#1}%
\providecommand \enquote  [1]{``#1''}%
\providecommand \bibnamefont  [1]{#1}%
\providecommand \bibfnamefont [1]{#1}%
\providecommand \citenamefont [1]{#1}%
\providecommand \href@noop [0]{\@secondoftwo}%
\providecommand \href [0]{\begingroup \@sanitize@url \@href}%
\providecommand \@href[1]{\@@startlink{#1}\@@href}%
\providecommand \@@href[1]{\endgroup#1\@@endlink}%
\providecommand \@sanitize@url [0]{\catcode `\\12\catcode `\$12\catcode
  `\&12\catcode `\#12\catcode `\^12\catcode `\_12\catcode `\%12\relax}%
\providecommand \@@startlink[1]{}%
\providecommand \@@endlink[0]{}%
\providecommand \url  [0]{\begingroup\@sanitize@url \@url }%
\providecommand \@url [1]{\endgroup\@href {#1}{\urlprefix }}%
\providecommand \urlprefix  [0]{URL }%
\providecommand \Eprint [0]{\href }%
\providecommand \doibase [0]{http://dx.doi.org/}%
\providecommand \selectlanguage [0]{\@gobble}%
\providecommand \bibinfo  [0]{\@secondoftwo}%
\providecommand \bibfield  [0]{\@secondoftwo}%
\providecommand \translation [1]{[#1]}%
\providecommand \BibitemOpen [0]{}%
\providecommand \bibitemStop [0]{}%
\providecommand \bibitemNoStop [0]{.\EOS\space}%
\providecommand \EOS [0]{\spacefactor3000\relax}%
\providecommand \BibitemShut  [1]{\csname bibitem#1\endcsname}%
\let\auto@bib@innerbib\@empty
%</preamble>
\bibitem [{\citenamefont {Aad}\ \emph {et~al.}(2012)\citenamefont {Aad} \emph
  {et~al.}}]{Aad:2012tfa}%
  \BibitemOpen
  \bibfield  {author} {\bibinfo {author} {\bibfnamefont {G.}~\bibnamefont
  {Aad}} \emph {et~al.} (\bibinfo {collaboration} {ATLAS}),\ }\href {\doibase
  10.1016/j.physletb.2012.08.020} {\bibfield  {journal} {\bibinfo  {journal}
  {Phys. Lett.}\ }\textbf {\bibinfo {volume} {B716}},\ \bibinfo {pages} {1}
  (\bibinfo {year} {2012})},\ \Eprint {http://arxiv.org/abs/1207.7214}
  {arXiv:1207.7214 [hep-ex]} \BibitemShut {NoStop}%
%%CITATION = ARXIV:1207.7214;%%
\bibitem [{\citenamefont {Chatrchyan}\ \emph {et~al.}(2012)\citenamefont
  {Chatrchyan} \emph {et~al.}}]{Chatrchyan:2012xdj}%
  \BibitemOpen
  \bibfield  {author} {\bibinfo {author} {\bibfnamefont {S.}~\bibnamefont
  {Chatrchyan}} \emph {et~al.} (\bibinfo {collaboration} {CMS}),\ }\href
  {\doibase 10.1016/j.physletb.2012.08.021} {\bibfield  {journal} {\bibinfo
  {journal} {Phys. Lett.}\ }\textbf {\bibinfo {volume} {B716}},\ \bibinfo
  {pages} {30} (\bibinfo {year} {2012})},\ \Eprint
  {http://arxiv.org/abs/1207.7235} {arXiv:1207.7235 [hep-ex]} \BibitemShut
  {NoStop}%
%%CITATION = ARXIV:1207.7235;%%
\bibitem [{ATL()}]{ATLASResults}%
  \BibitemOpen
  \href@noop {} {}\bibinfo {note} {See, for example,
  \url{https://atlas.web.cern.ch/Atlas/GROUPS/PHYSICS/PUBNOTES/ATL-PHYS-PUB-2020-013/}
  and
  \url{https://atlas.web.cern.ch/Atlas/GROUPS/PHYSICS/CombinedSummaryPlots/EXOTICS/}}\BibitemShut
  {NoStop}%
\bibitem [{CMS()}]{CMSResults}%
  \BibitemOpen
  \href@noop {} {}\bibinfo {note} {See, for example,
  \url{https://twiki.cern.ch/twiki/bin/view/CMSPublic/PhysicsResultsSUS} and
  \url{https://twiki.cern.ch/twiki/bin/view/CMSPublic/PhysicsResultsB2G}}\BibitemShut
  {NoStop}%
\bibitem [{\citenamefont {Chacko}\ \emph {et~al.}(2006)\citenamefont {Chacko},
  \citenamefont {Goh},\ and\ \citenamefont {Harnik}}]{Chacko:2005pe}%
  \BibitemOpen
  \bibfield  {author} {\bibinfo {author} {\bibfnamefont {Z.}~\bibnamefont
  {Chacko}}, \bibinfo {author} {\bibfnamefont {H.-S.}\ \bibnamefont {Goh}}, \
  and\ \bibinfo {author} {\bibfnamefont {R.}~\bibnamefont {Harnik}},\ }\href
  {\doibase 10.1103/PhysRevLett.96.231802} {\bibfield  {journal} {\bibinfo
  {journal} {Phys. Rev. Lett.}\ }\textbf {\bibinfo {volume} {96}},\ \bibinfo
  {pages} {231802} (\bibinfo {year} {2006})},\ \Eprint
  {http://arxiv.org/abs/hep-ph/0506256} {arXiv:hep-ph/0506256} \BibitemShut
  {NoStop}%
\bibitem [{\citenamefont {Graham}\ \emph {et~al.}(2015)\citenamefont {Graham},
  \citenamefont {Kaplan},\ and\ \citenamefont {Rajendran}}]{Graham:2015cka}%
  \BibitemOpen
  \bibfield  {author} {\bibinfo {author} {\bibfnamefont {P.~W.}\ \bibnamefont
  {Graham}}, \bibinfo {author} {\bibfnamefont {D.~E.}\ \bibnamefont {Kaplan}},
  \ and\ \bibinfo {author} {\bibfnamefont {S.}~\bibnamefont {Rajendran}},\
  }\href {\doibase 10.1103/PhysRevLett.115.221801} {\bibfield  {journal}
  {\bibinfo  {journal} {Phys. Rev. Lett.}\ }\textbf {\bibinfo {volume} {115}},\
  \bibinfo {pages} {221801} (\bibinfo {year} {2015})},\ \Eprint
  {http://arxiv.org/abs/1504.07551} {arXiv:1504.07551 [hep-ph]} \BibitemShut
  {NoStop}%
\bibitem [{\citenamefont {Arkani-Hamed}\ \emph {et~al.}(2016)\citenamefont
  {Arkani-Hamed}, \citenamefont {Cohen}, \citenamefont {D'Agnolo},
  \citenamefont {Hook}, \citenamefont {Kim},\ and\ \citenamefont
  {Pinner}}]{Arkani-Hamed:2016rle}%
  \BibitemOpen
  \bibfield  {author} {\bibinfo {author} {\bibfnamefont {N.}~\bibnamefont
  {Arkani-Hamed}}, \bibinfo {author} {\bibfnamefont {T.}~\bibnamefont {Cohen}},
  \bibinfo {author} {\bibfnamefont {R.~T.}\ \bibnamefont {D'Agnolo}}, \bibinfo
  {author} {\bibfnamefont {A.}~\bibnamefont {Hook}}, \bibinfo {author}
  {\bibfnamefont {H.~D.}\ \bibnamefont {Kim}}, \ and\ \bibinfo {author}
  {\bibfnamefont {D.}~\bibnamefont {Pinner}},\ }\href {\doibase
  10.1103/PhysRevLett.117.251801} {\bibfield  {journal} {\bibinfo  {journal}
  {Phys. Rev. Lett.}\ }\textbf {\bibinfo {volume} {117}},\ \bibinfo {pages}
  {251801} (\bibinfo {year} {2016})},\ \Eprint
  {http://arxiv.org/abs/1607.06821} {arXiv:1607.06821 [hep-ph]} \BibitemShut
  {NoStop}%
%%CITATION = ARXIV:1607.06821;%%
\bibitem [{\citenamefont {Geller}\ \emph {et~al.}(2019)\citenamefont {Geller},
  \citenamefont {Hochberg},\ and\ \citenamefont {Kuflik}}]{Geller:2018xvz}%
  \BibitemOpen
  \bibfield  {author} {\bibinfo {author} {\bibfnamefont {M.}~\bibnamefont
  {Geller}}, \bibinfo {author} {\bibfnamefont {Y.}~\bibnamefont {Hochberg}}, \
  and\ \bibinfo {author} {\bibfnamefont {E.}~\bibnamefont {Kuflik}},\ }\href
  {\doibase 10.1103/PhysRevLett.122.191802} {\bibfield  {journal} {\bibinfo
  {journal} {Phys. Rev. Lett.}\ }\textbf {\bibinfo {volume} {122}},\ \bibinfo
  {pages} {191802} (\bibinfo {year} {2019})},\ \Eprint
  {http://arxiv.org/abs/1809.07338} {arXiv:1809.07338 [hep-ph]} \BibitemShut
  {NoStop}%
%%CITATION = ARXIV:1809.07338;%%
\bibitem [{\citenamefont {Cheung}\ and\ \citenamefont
  {Saraswat}(2018)}]{Cheung:2018xnu}%
  \BibitemOpen
  \bibfield  {author} {\bibinfo {author} {\bibfnamefont {C.}~\bibnamefont
  {Cheung}}\ and\ \bibinfo {author} {\bibfnamefont {P.}~\bibnamefont
  {Saraswat}},\ }\href@noop {} {\  (\bibinfo {year} {2018})},\ \Eprint
  {http://arxiv.org/abs/1811.12390} {arXiv:1811.12390 [hep-ph]} \BibitemShut
  {NoStop}%
%%CITATION = ARXIV:1811.12390;%%
\bibitem [{\citenamefont {Strumia}\ and\ \citenamefont
  {Teresi}(2020)}]{Strumia:2020bdy}%
  \BibitemOpen
  \bibfield  {author} {\bibinfo {author} {\bibfnamefont {A.}~\bibnamefont
  {Strumia}}\ and\ \bibinfo {author} {\bibfnamefont {D.}~\bibnamefont
  {Teresi}},\ }\href@noop {} {\  (\bibinfo {year} {2020})},\ \Eprint
  {http://arxiv.org/abs/2002.02463} {arXiv:2002.02463 [hep-ph]} \BibitemShut
  {NoStop}%
%%CITATION = ARXIV:2002.02463;%%
\bibitem [{\citenamefont {Giudice}\ \emph {et~al.}(2019)\citenamefont
  {Giudice}, \citenamefont {Kehagias},\ and\ \citenamefont
  {Riotto}}]{Giudice:2019iwl}%
  \BibitemOpen
  \bibfield  {author} {\bibinfo {author} {\bibfnamefont {G.~F.}\ \bibnamefont
  {Giudice}}, \bibinfo {author} {\bibfnamefont {A.}~\bibnamefont {Kehagias}}, \
  and\ \bibinfo {author} {\bibfnamefont {A.}~\bibnamefont {Riotto}},\ }\href
  {\doibase 10.1007/JHEP10(2019)199} {\bibfield  {journal} {\bibinfo  {journal}
  {JHEP}\ }\textbf {\bibinfo {volume} {10}},\ \bibinfo {pages} {199} (\bibinfo
  {year} {2019})},\ \Eprint {http://arxiv.org/abs/1907.05370} {arXiv:1907.05370
  [hep-ph]} \BibitemShut {NoStop}%
%%CITATION = ARXIV:1907.05370;%%
\bibitem [{\citenamefont {Agrawal}\ \emph {et~al.}(1998)\citenamefont
  {Agrawal}, \citenamefont {Barr}, \citenamefont {Donoghue},\ and\
  \citenamefont {Seckel}}]{Agrawal:1997gf}%
  \BibitemOpen
  \bibfield  {author} {\bibinfo {author} {\bibfnamefont {V.}~\bibnamefont
  {Agrawal}}, \bibinfo {author} {\bibfnamefont {S.~M.}\ \bibnamefont {Barr}},
  \bibinfo {author} {\bibfnamefont {J.~F.}\ \bibnamefont {Donoghue}}, \ and\
  \bibinfo {author} {\bibfnamefont {D.}~\bibnamefont {Seckel}},\ }\href
  {\doibase 10.1103/PhysRevD.57.5480} {\bibfield  {journal} {\bibinfo
  {journal} {Phys. Rev. D}\ }\textbf {\bibinfo {volume} {57}},\ \bibinfo
  {pages} {5480} (\bibinfo {year} {1998})},\ \Eprint
  {http://arxiv.org/abs/hep-ph/9707380} {arXiv:hep-ph/9707380} \BibitemShut
  {NoStop}%
\bibitem [{\citenamefont {Arkani-Hamed}\ and\ \citenamefont
  {Dimopoulos}(2005)}]{ArkaniHamed:2004fb}%
  \BibitemOpen
  \bibfield  {author} {\bibinfo {author} {\bibfnamefont {N.}~\bibnamefont
  {Arkani-Hamed}}\ and\ \bibinfo {author} {\bibfnamefont {S.}~\bibnamefont
  {Dimopoulos}},\ }\href {\doibase 10.1088/1126-6708/2005/06/073} {\bibfield
  {journal} {\bibinfo  {journal} {JHEP}\ }\textbf {\bibinfo {volume} {06}},\
  \bibinfo {pages} {073} (\bibinfo {year} {2005})},\ \Eprint
  {http://arxiv.org/abs/hep-th/0405159} {arXiv:hep-th/0405159} \BibitemShut
  {NoStop}%
\bibitem [{\citenamefont {Randall}\ and\ \citenamefont
  {Sundrum}(1999)}]{Randall:1999ee}%
  \BibitemOpen
  \bibfield  {author} {\bibinfo {author} {\bibfnamefont {L.}~\bibnamefont
  {Randall}}\ and\ \bibinfo {author} {\bibfnamefont {R.}~\bibnamefont
  {Sundrum}},\ }\href {\doibase 10.1103/PhysRevLett.83.3370} {\bibfield
  {journal} {\bibinfo  {journal} {Phys. Rev. Lett.}\ }\textbf {\bibinfo
  {volume} {83}},\ \bibinfo {pages} {3370} (\bibinfo {year} {1999})},\ \Eprint
  {http://arxiv.org/abs/hep-ph/9905221} {arXiv:hep-ph/9905221 [hep-ph]}
  \BibitemShut {NoStop}%
%%CITATION = HEP-PH/9905221;%%
\bibitem [{\citenamefont {Csaki}\ \emph {et~al.}(2000)\citenamefont {Csaki},
  \citenamefont {Graesser}, \citenamefont {Randall},\ and\ \citenamefont
  {Terning}}]{Csaki:1999mp}%
  \BibitemOpen
  \bibfield  {author} {\bibinfo {author} {\bibfnamefont {C.}~\bibnamefont
  {Csaki}}, \bibinfo {author} {\bibfnamefont {M.}~\bibnamefont {Graesser}},
  \bibinfo {author} {\bibfnamefont {L.}~\bibnamefont {Randall}}, \ and\
  \bibinfo {author} {\bibfnamefont {J.}~\bibnamefont {Terning}},\ }\href
  {\doibase 10.1103/PhysRevD.62.045015} {\bibfield  {journal} {\bibinfo
  {journal} {Phys. Rev.}\ }\textbf {\bibinfo {volume} {D62}},\ \bibinfo {pages}
  {045015} (\bibinfo {year} {2000})},\ \Eprint
  {http://arxiv.org/abs/hep-ph/9911406} {arXiv:hep-ph/9911406 [hep-ph]}
  \BibitemShut {NoStop}%
%%CITATION = HEP-PH/9911406;%%
\bibitem [{\citenamefont {Csaki}\ \emph {et~al.}(2001)\citenamefont {Csaki},
  \citenamefont {Graesser},\ and\ \citenamefont {Kribs}}]{Csaki:2000zn}%
  \BibitemOpen
  \bibfield  {author} {\bibinfo {author} {\bibfnamefont {C.}~\bibnamefont
  {Csaki}}, \bibinfo {author} {\bibfnamefont {M.~L.}\ \bibnamefont {Graesser}},
  \ and\ \bibinfo {author} {\bibfnamefont {G.~D.}\ \bibnamefont {Kribs}},\
  }\href {\doibase 10.1103/PhysRevD.63.065002} {\bibfield  {journal} {\bibinfo
  {journal} {Phys. Rev.}\ }\textbf {\bibinfo {volume} {D63}},\ \bibinfo {pages}
  {065002} (\bibinfo {year} {2001})},\ \Eprint
  {http://arxiv.org/abs/hep-th/0008151} {arXiv:hep-th/0008151 [hep-th]}
  \BibitemShut {NoStop}%
%%CITATION = HEP-TH/0008151;%%
\bibitem [{\citenamefont {Rattazzi}\ and\ \citenamefont
  {Zaffaroni}(2001)}]{Rattazzi:2000hs}%
  \BibitemOpen
  \bibfield  {author} {\bibinfo {author} {\bibfnamefont {R.}~\bibnamefont
  {Rattazzi}}\ and\ \bibinfo {author} {\bibfnamefont {A.}~\bibnamefont
  {Zaffaroni}},\ }\href {\doibase 10.1088/1126-6708/2001/04/021} {\bibfield
  {journal} {\bibinfo  {journal} {JHEP}\ }\textbf {\bibinfo {volume} {04}},\
  \bibinfo {pages} {021} (\bibinfo {year} {2001})},\ \Eprint
  {http://arxiv.org/abs/hep-th/0012248} {arXiv:hep-th/0012248 [hep-th]}
  \BibitemShut {NoStop}%
%%CITATION = HEP-TH/0012248;%%
\bibitem [{\citenamefont {Goldberger}\ and\ \citenamefont
  {Wise}(1999)}]{Goldberger:1999uk}%
  \BibitemOpen
  \bibfield  {author} {\bibinfo {author} {\bibfnamefont {W.~D.}\ \bibnamefont
  {Goldberger}}\ and\ \bibinfo {author} {\bibfnamefont {M.~B.}\ \bibnamefont
  {Wise}},\ }\href {\doibase 10.1103/PhysRevLett.83.4922} {\bibfield  {journal}
  {\bibinfo  {journal} {Phys. Rev. Lett.}\ }\textbf {\bibinfo {volume} {83}},\
  \bibinfo {pages} {4922} (\bibinfo {year} {1999})},\ \Eprint
  {http://arxiv.org/abs/hep-ph/9907447} {arXiv:hep-ph/9907447} \BibitemShut
  {NoStop}%
\bibitem [{\citenamefont {Gupta}\ \emph {et~al.}(2016)\citenamefont {Gupta},
  \citenamefont {Komargodski}, \citenamefont {Perez},\ and\ \citenamefont
  {Ubaldi}}]{Gupta:2015uea}%
  \BibitemOpen
  \bibfield  {author} {\bibinfo {author} {\bibfnamefont {R.~S.}\ \bibnamefont
  {Gupta}}, \bibinfo {author} {\bibfnamefont {Z.}~\bibnamefont {Komargodski}},
  \bibinfo {author} {\bibfnamefont {G.}~\bibnamefont {Perez}}, \ and\ \bibinfo
  {author} {\bibfnamefont {L.}~\bibnamefont {Ubaldi}},\ }\href {\doibase
  10.1007/JHEP02(2016)166} {\bibfield  {journal} {\bibinfo  {journal} {JHEP}\
  }\textbf {\bibinfo {volume} {02}},\ \bibinfo {pages} {166} (\bibinfo {year}
  {2016})},\ \Eprint {http://arxiv.org/abs/1509.00047} {arXiv:1509.00047
  [hep-ph]} \BibitemShut {NoStop}%
%%CITATION = ARXIV:1509.00047;%%
\bibitem [{\citenamefont {Csaki}\ \emph {et~al.}(2007)\citenamefont {Csaki},
  \citenamefont {Hubisz},\ and\ \citenamefont {Lee}}]{Csaki:2007ns}%
  \BibitemOpen
  \bibfield  {author} {\bibinfo {author} {\bibfnamefont {C.}~\bibnamefont
  {Csaki}}, \bibinfo {author} {\bibfnamefont {J.}~\bibnamefont {Hubisz}}, \
  and\ \bibinfo {author} {\bibfnamefont {S.~J.}\ \bibnamefont {Lee}},\ }\href
  {\doibase 10.1103/PhysRevD.76.125015} {\bibfield  {journal} {\bibinfo
  {journal} {Phys. Rev. D}\ }\textbf {\bibinfo {volume} {76}},\ \bibinfo
  {pages} {125015} (\bibinfo {year} {2007})},\ \Eprint
  {http://arxiv.org/abs/0705.3844} {arXiv:0705.3844 [hep-ph]} \BibitemShut
  {NoStop}%
\bibitem [{\citenamefont {Coleman}(1977)}]{Coleman:1977py}%
  \BibitemOpen
  \bibfield  {author} {\bibinfo {author} {\bibfnamefont {S.~R.}\ \bibnamefont
  {Coleman}},\ }\href {\doibase 10.1103/PhysRevD.16.1248} {\bibfield  {journal}
  {\bibinfo  {journal} {Phys. Rev. D}\ }\textbf {\bibinfo {volume} {15}},\
  \bibinfo {pages} {2929} (\bibinfo {year} {1977})},\ \bibinfo {note}
  {[Erratum: Phys.Rev.D 16, 1248 (1977)]}\BibitemShut {NoStop}%
\bibitem [{\citenamefont {Aaij}\ \emph {et~al.}(2013)\citenamefont {Aaij} \emph
  {et~al.}}]{Aaij:2012vr}%
  \BibitemOpen
  \bibfield  {author} {\bibinfo {author} {\bibfnamefont {R.}~\bibnamefont
  {Aaij}} \emph {et~al.} (\bibinfo {collaboration} {LHCb}),\ }\href {\doibase
  10.1007/JHEP02(2013)105} {\bibfield  {journal} {\bibinfo  {journal} {JHEP}\
  }\textbf {\bibinfo {volume} {02}},\ \bibinfo {pages} {105} (\bibinfo {year}
  {2013})},\ \Eprint {http://arxiv.org/abs/1209.4284} {arXiv:1209.4284
  [hep-ex]} \BibitemShut {NoStop}%
\bibitem [{\citenamefont {Aaij}\ \emph {et~al.}(2015)\citenamefont {Aaij} \emph
  {et~al.}}]{Aaij:2015tna}%
  \BibitemOpen
  \bibfield  {author} {\bibinfo {author} {\bibfnamefont {R.}~\bibnamefont
  {Aaij}} \emph {et~al.} (\bibinfo {collaboration} {LHCb}),\ }\href {\doibase
  10.1103/PhysRevLett.115.161802} {\bibfield  {journal} {\bibinfo  {journal}
  {Phys. Rev. Lett.}\ }\textbf {\bibinfo {volume} {115}},\ \bibinfo {pages}
  {161802} (\bibinfo {year} {2015})},\ \Eprint
  {http://arxiv.org/abs/1508.04094} {arXiv:1508.04094 [hep-ex]} \BibitemShut
  {NoStop}%
\bibitem [{\citenamefont {Flacke}\ \emph {et~al.}(2017)\citenamefont {Flacke},
  \citenamefont {Frugiuele}, \citenamefont {Fuchs}, \citenamefont {Gupta},\
  and\ \citenamefont {Perez}}]{Flacke:2016szy}%
  \BibitemOpen
  \bibfield  {author} {\bibinfo {author} {\bibfnamefont {T.}~\bibnamefont
  {Flacke}}, \bibinfo {author} {\bibfnamefont {C.}~\bibnamefont {Frugiuele}},
  \bibinfo {author} {\bibfnamefont {E.}~\bibnamefont {Fuchs}}, \bibinfo
  {author} {\bibfnamefont {R.~S.}\ \bibnamefont {Gupta}}, \ and\ \bibinfo
  {author} {\bibfnamefont {G.}~\bibnamefont {Perez}},\ }\href {\doibase
  10.1007/JHEP06(2017)050} {\bibfield  {journal} {\bibinfo  {journal} {JHEP}\
  }\textbf {\bibinfo {volume} {06}},\ \bibinfo {pages} {050} (\bibinfo {year}
  {2017})},\ \Eprint {http://arxiv.org/abs/1610.02025} {arXiv:1610.02025
  [hep-ph]} \BibitemShut {NoStop}%
\bibitem [{\citenamefont {Frugiuele}\ \emph {et~al.}(2018)\citenamefont
  {Frugiuele}, \citenamefont {Fuchs}, \citenamefont {Perez},\ and\
  \citenamefont {Schlaffer}}]{Frugiuele:2018coc}%
  \BibitemOpen
  \bibfield  {author} {\bibinfo {author} {\bibfnamefont {C.}~\bibnamefont
  {Frugiuele}}, \bibinfo {author} {\bibfnamefont {E.}~\bibnamefont {Fuchs}},
  \bibinfo {author} {\bibfnamefont {G.}~\bibnamefont {Perez}}, \ and\ \bibinfo
  {author} {\bibfnamefont {M.}~\bibnamefont {Schlaffer}},\ }\href {\doibase
  10.1007/JHEP10(2018)151} {\bibfield  {journal} {\bibinfo  {journal} {JHEP}\
  }\textbf {\bibinfo {volume} {10}},\ \bibinfo {pages} {151} (\bibinfo {year}
  {2018})},\ \Eprint {http://arxiv.org/abs/1807.10842} {arXiv:1807.10842
  [hep-ph]} \BibitemShut {NoStop}%
\bibitem [{\citenamefont {Alekhin}\ \emph {et~al.}(2016)\citenamefont {Alekhin}
  \emph {et~al.}}]{Alekhin:2015byh}%
  \BibitemOpen
  \bibfield  {author} {\bibinfo {author} {\bibfnamefont {S.}~\bibnamefont
  {Alekhin}} \emph {et~al.},\ }\href {\doibase 10.1088/0034-4885/79/12/124201}
  {\bibfield  {journal} {\bibinfo  {journal} {Rept. Prog. Phys.}\ }\textbf
  {\bibinfo {volume} {79}},\ \bibinfo {pages} {124201} (\bibinfo {year}
  {2016})},\ \Eprint {http://arxiv.org/abs/1504.04855} {arXiv:1504.04855
  [hep-ph]} \BibitemShut {NoStop}%
\bibitem [{\citenamefont {Gardner}\ \emph {et~al.}(2016)\citenamefont
  {Gardner}, \citenamefont {Holt},\ and\ \citenamefont
  {Tadepalli}}]{Gardner:2015wea}%
  \BibitemOpen
  \bibfield  {author} {\bibinfo {author} {\bibfnamefont {S.}~\bibnamefont
  {Gardner}}, \bibinfo {author} {\bibfnamefont {R.}~\bibnamefont {Holt}}, \
  and\ \bibinfo {author} {\bibfnamefont {A.}~\bibnamefont {Tadepalli}},\ }\href
  {\doibase 10.1103/PhysRevD.93.115015} {\bibfield  {journal} {\bibinfo
  {journal} {Phys. Rev. D}\ }\textbf {\bibinfo {volume} {93}},\ \bibinfo
  {pages} {115015} (\bibinfo {year} {2016})},\ \Eprint
  {http://arxiv.org/abs/1509.00050} {arXiv:1509.00050 [hep-ph]} \BibitemShut
  {NoStop}%
\bibitem [{\citenamefont {Beacham}\ \emph {et~al.}(2020)\citenamefont {Beacham}
  \emph {et~al.}}]{Beacham:2019nyx}%
  \BibitemOpen
  \bibfield  {author} {\bibinfo {author} {\bibfnamefont {J.}~\bibnamefont
  {Beacham}} \emph {et~al.},\ }\href {\doibase 10.1088/1361-6471/ab4cd2}
  {\bibfield  {journal} {\bibinfo  {journal} {J. Phys. G}\ }\textbf {\bibinfo
  {volume} {47}},\ \bibinfo {pages} {010501} (\bibinfo {year} {2020})},\
  \Eprint {http://arxiv.org/abs/1901.09966} {arXiv:1901.09966 [hep-ex]}
  \BibitemShut {NoStop}%
\bibitem [{\citenamefont {Feng}\ \emph
  {et~al.}(2018{\natexlab{a}})\citenamefont {Feng}, \citenamefont {Galon},
  \citenamefont {Kling},\ and\ \citenamefont {Trojanowski}}]{Feng:2017uoz}%
  \BibitemOpen
  \bibfield  {author} {\bibinfo {author} {\bibfnamefont {J.~L.}\ \bibnamefont
  {Feng}}, \bibinfo {author} {\bibfnamefont {I.}~\bibnamefont {Galon}},
  \bibinfo {author} {\bibfnamefont {F.}~\bibnamefont {Kling}}, \ and\ \bibinfo
  {author} {\bibfnamefont {S.}~\bibnamefont {Trojanowski}},\ }\href {\doibase
  10.1103/PhysRevD.97.035001} {\bibfield  {journal} {\bibinfo  {journal} {Phys.
  Rev. D}\ }\textbf {\bibinfo {volume} {97}},\ \bibinfo {pages} {035001}
  (\bibinfo {year} {2018}{\natexlab{a}})},\ \Eprint
  {http://arxiv.org/abs/1708.09389} {arXiv:1708.09389 [hep-ph]} \BibitemShut
  {NoStop}%
\bibitem [{\citenamefont {Feng}\ \emph
  {et~al.}(2018{\natexlab{b}})\citenamefont {Feng}, \citenamefont {Galon},
  \citenamefont {Kling},\ and\ \citenamefont {Trojanowski}}]{Feng:2017vli}%
  \BibitemOpen
  \bibfield  {author} {\bibinfo {author} {\bibfnamefont {J.~L.}\ \bibnamefont
  {Feng}}, \bibinfo {author} {\bibfnamefont {I.}~\bibnamefont {Galon}},
  \bibinfo {author} {\bibfnamefont {F.}~\bibnamefont {Kling}}, \ and\ \bibinfo
  {author} {\bibfnamefont {S.}~\bibnamefont {Trojanowski}},\ }\href {\doibase
  10.1103/PhysRevD.97.055034} {\bibfield  {journal} {\bibinfo  {journal} {Phys.
  Rev. D}\ }\textbf {\bibinfo {volume} {97}},\ \bibinfo {pages} {055034}
  (\bibinfo {year} {2018}{\natexlab{b}})},\ \Eprint
  {http://arxiv.org/abs/1710.09387} {arXiv:1710.09387 [hep-ph]} \BibitemShut
  {NoStop}%
\bibitem [{\citenamefont {Gligorov}\ \emph {et~al.}(2018)\citenamefont
  {Gligorov}, \citenamefont {Knapen}, \citenamefont {Papucci},\ and\
  \citenamefont {Robinson}}]{Gligorov:2017nwh}%
  \BibitemOpen
  \bibfield  {author} {\bibinfo {author} {\bibfnamefont {V.~V.}\ \bibnamefont
  {Gligorov}}, \bibinfo {author} {\bibfnamefont {S.}~\bibnamefont {Knapen}},
  \bibinfo {author} {\bibfnamefont {M.}~\bibnamefont {Papucci}}, \ and\
  \bibinfo {author} {\bibfnamefont {D.~J.}\ \bibnamefont {Robinson}},\ }\href
  {\doibase 10.1103/PhysRevD.97.015023} {\bibfield  {journal} {\bibinfo
  {journal} {Phys. Rev. D}\ }\textbf {\bibinfo {volume} {97}},\ \bibinfo
  {pages} {015023} (\bibinfo {year} {2018})},\ \Eprint
  {http://arxiv.org/abs/1708.09395} {arXiv:1708.09395 [hep-ph]} \BibitemShut
  {NoStop}%
\bibitem [{\citenamefont {Evans}\ \emph {et~al.}(2018)\citenamefont {Evans},
  \citenamefont {Gori},\ and\ \citenamefont {Shelton}}]{Evans:2017kti}%
  \BibitemOpen
  \bibfield  {author} {\bibinfo {author} {\bibfnamefont {J.~A.}\ \bibnamefont
  {Evans}}, \bibinfo {author} {\bibfnamefont {S.}~\bibnamefont {Gori}}, \ and\
  \bibinfo {author} {\bibfnamefont {J.}~\bibnamefont {Shelton}},\ }\href
  {\doibase 10.1007/JHEP02(2018)100} {\bibfield  {journal} {\bibinfo  {journal}
  {JHEP}\ }\textbf {\bibinfo {volume} {02}},\ \bibinfo {pages} {100} (\bibinfo
  {year} {2018})},\ \Eprint {http://arxiv.org/abs/1712.03974} {arXiv:1712.03974
  [hep-ph]} \BibitemShut {NoStop}%
\bibitem [{\citenamefont {Chou}\ \emph {et~al.}(2017)\citenamefont {Chou},
  \citenamefont {Curtin},\ and\ \citenamefont {Lubatti}}]{Chou:2016lxi}%
  \BibitemOpen
  \bibfield  {author} {\bibinfo {author} {\bibfnamefont {J.~P.}\ \bibnamefont
  {Chou}}, \bibinfo {author} {\bibfnamefont {D.}~\bibnamefont {Curtin}}, \ and\
  \bibinfo {author} {\bibfnamefont {H.}~\bibnamefont {Lubatti}},\ }\href
  {\doibase 10.1016/j.physletb.2017.01.043} {\bibfield  {journal} {\bibinfo
  {journal} {Phys. Lett. B}\ }\textbf {\bibinfo {volume} {767}},\ \bibinfo
  {pages} {29} (\bibinfo {year} {2017})},\ \Eprint
  {http://arxiv.org/abs/1606.06298} {arXiv:1606.06298 [hep-ph]} \BibitemShut
  {NoStop}%
\bibitem [{\citenamefont {Evans}(2018)}]{Evans:2017lvd}%
  \BibitemOpen
  \bibfield  {author} {\bibinfo {author} {\bibfnamefont {J.~A.}\ \bibnamefont
  {Evans}},\ }\href {\doibase 10.1103/PhysRevD.97.055046} {\bibfield  {journal}
  {\bibinfo  {journal} {Phys. Rev. D}\ }\textbf {\bibinfo {volume} {97}},\
  \bibinfo {pages} {055046} (\bibinfo {year} {2018})},\ \Eprint
  {http://arxiv.org/abs/1708.08503} {arXiv:1708.08503 [hep-ph]} \BibitemShut
  {NoStop}%
\bibitem [{\citenamefont {Acciarri}\ \emph {et~al.}(1996)\citenamefont
  {Acciarri} \emph {et~al.}}]{Acciarri:1996um}%
  \BibitemOpen
  \bibfield  {author} {\bibinfo {author} {\bibfnamefont {M.}~\bibnamefont
  {Acciarri}} \emph {et~al.} (\bibinfo {collaboration} {L3}),\ }\href {\doibase
  10.1016/0370-2693(96)00987-2} {\bibfield  {journal} {\bibinfo  {journal}
  {Phys. Lett. B}\ }\textbf {\bibinfo {volume} {385}},\ \bibinfo {pages} {454}
  (\bibinfo {year} {1996})}\BibitemShut {NoStop}%
\bibitem [{\citenamefont {Schael}\ \emph {et~al.}(2006)\citenamefont {Schael}
  \emph {et~al.}}]{Schael:2006cr}%
  \BibitemOpen
  \bibfield  {author} {\bibinfo {author} {\bibfnamefont {S.}~\bibnamefont
  {Schael}} \emph {et~al.} (\bibinfo {collaboration} {ALEPH, DELPHI, L3, OPAL,
  LEP Working Group for Higgs Boson Searches}),\ }\href {\doibase
  10.1140/epjc/s2006-02569-7} {\bibfield  {journal} {\bibinfo  {journal} {Eur.
  Phys. J. C}\ }\textbf {\bibinfo {volume} {47}},\ \bibinfo {pages} {547}
  (\bibinfo {year} {2006})},\ \Eprint {http://arxiv.org/abs/hep-ex/0602042}
  {arXiv:hep-ex/0602042} \BibitemShut {NoStop}%
\bibitem [{\citenamefont {Abada}\ \emph {et~al.}(2019)\citenamefont {Abada}
  \emph {et~al.}}]{Abada:2019zxq}%
  \BibitemOpen
  \bibfield  {author} {\bibinfo {author} {\bibfnamefont {A.}~\bibnamefont
  {Abada}} \emph {et~al.} (\bibinfo {collaboration} {FCC}),\ }\href {\doibase
  10.1140/epjst/e2019-900045-4} {\bibfield  {journal} {\bibinfo  {journal}
  {Eur. Phys. J. ST}\ }\textbf {\bibinfo {volume} {228}},\ \bibinfo {pages}
  {261} (\bibinfo {year} {2019})}\BibitemShut {NoStop}%
\bibitem [{\citenamefont {Bauer}\ \emph {et~al.}(2019)\citenamefont {Bauer},
  \citenamefont {Heiles}, \citenamefont {Neubert},\ and\ \citenamefont
  {Thamm}}]{Bauer:2018uxu}%
  \BibitemOpen
  \bibfield  {author} {\bibinfo {author} {\bibfnamefont {M.}~\bibnamefont
  {Bauer}}, \bibinfo {author} {\bibfnamefont {M.}~\bibnamefont {Heiles}},
  \bibinfo {author} {\bibfnamefont {M.}~\bibnamefont {Neubert}}, \ and\
  \bibinfo {author} {\bibfnamefont {A.}~\bibnamefont {Thamm}},\ }\href
  {\doibase 10.1140/epjc/s10052-019-6587-9} {\bibfield  {journal} {\bibinfo
  {journal} {Eur. Phys. J. C}\ }\textbf {\bibinfo {volume} {79}},\ \bibinfo
  {pages} {74} (\bibinfo {year} {2019})},\ \Eprint
  {http://arxiv.org/abs/1808.10323} {arXiv:1808.10323 [hep-ph]} \BibitemShut
  {NoStop}%
\bibitem [{\citenamefont {Mimasu}\ and\ \citenamefont
  {Sanz}(2015)}]{Mimasu:2014nea}%
  \BibitemOpen
  \bibfield  {author} {\bibinfo {author} {\bibfnamefont {K.}~\bibnamefont
  {Mimasu}}\ and\ \bibinfo {author} {\bibfnamefont {V.}~\bibnamefont {Sanz}},\
  }\href {\doibase 10.1007/JHEP06(2015)173} {\bibfield  {journal} {\bibinfo
  {journal} {JHEP}\ }\textbf {\bibinfo {volume} {06}},\ \bibinfo {pages} {173}
  (\bibinfo {year} {2015})},\ \Eprint {http://arxiv.org/abs/1409.4792}
  {arXiv:1409.4792 [hep-ph]} \BibitemShut {NoStop}%
\bibitem [{\citenamefont {Jaeckel}\ and\ \citenamefont
  {Spannowsky}(2016)}]{Jaeckel:2015jla}%
  \BibitemOpen
  \bibfield  {author} {\bibinfo {author} {\bibfnamefont {J.}~\bibnamefont
  {Jaeckel}}\ and\ \bibinfo {author} {\bibfnamefont {M.}~\bibnamefont
  {Spannowsky}},\ }\href {\doibase 10.1016/j.physletb.2015.12.037} {\bibfield
  {journal} {\bibinfo  {journal} {Phys. Lett. B}\ }\textbf {\bibinfo {volume}
  {753}},\ \bibinfo {pages} {482} (\bibinfo {year} {2016})},\ \Eprint
  {http://arxiv.org/abs/1509.00476} {arXiv:1509.00476 [hep-ph]} \BibitemShut
  {NoStop}%
\bibitem [{\citenamefont {Bauer}\ \emph {et~al.}(2017)\citenamefont {Bauer},
  \citenamefont {Neubert},\ and\ \citenamefont {Thamm}}]{Bauer:2017ris}%
  \BibitemOpen
  \bibfield  {author} {\bibinfo {author} {\bibfnamefont {M.}~\bibnamefont
  {Bauer}}, \bibinfo {author} {\bibfnamefont {M.}~\bibnamefont {Neubert}}, \
  and\ \bibinfo {author} {\bibfnamefont {A.}~\bibnamefont {Thamm}},\ }\href
  {\doibase 10.1007/JHEP12(2017)044} {\bibfield  {journal} {\bibinfo  {journal}
  {JHEP}\ }\textbf {\bibinfo {volume} {12}},\ \bibinfo {pages} {044} (\bibinfo
  {year} {2017})},\ \Eprint {http://arxiv.org/abs/1708.00443} {arXiv:1708.00443
  [hep-ph]} \BibitemShut {NoStop}%
\bibitem [{\citenamefont {Knapen}\ \emph {et~al.}(2017)\citenamefont {Knapen},
  \citenamefont {Lin}, \citenamefont {Lou},\ and\ \citenamefont
  {Melia}}]{Knapen:2016moh}%
  \BibitemOpen
  \bibfield  {author} {\bibinfo {author} {\bibfnamefont {S.}~\bibnamefont
  {Knapen}}, \bibinfo {author} {\bibfnamefont {T.}~\bibnamefont {Lin}},
  \bibinfo {author} {\bibfnamefont {H.~K.}\ \bibnamefont {Lou}}, \ and\
  \bibinfo {author} {\bibfnamefont {T.}~\bibnamefont {Melia}},\ }\href
  {\doibase 10.1103/PhysRevLett.118.171801} {\bibfield  {journal} {\bibinfo
  {journal} {Phys. Rev. Lett.}\ }\textbf {\bibinfo {volume} {118}},\ \bibinfo
  {pages} {171801} (\bibinfo {year} {2017})},\ \Eprint
  {http://arxiv.org/abs/1607.06083} {arXiv:1607.06083 [hep-ph]} \BibitemShut
  {NoStop}%
\bibitem [{\citenamefont {Knapen}\ \emph {et~al.}(2018)\citenamefont {Knapen},
  \citenamefont {Lin}, \citenamefont {Lou},\ and\ \citenamefont
  {Melia}}]{Knapen:2017ebd}%
  \BibitemOpen
  \bibfield  {author} {\bibinfo {author} {\bibfnamefont {S.}~\bibnamefont
  {Knapen}}, \bibinfo {author} {\bibfnamefont {T.}~\bibnamefont {Lin}},
  \bibinfo {author} {\bibfnamefont {H.~K.}\ \bibnamefont {Lou}}, \ and\
  \bibinfo {author} {\bibfnamefont {T.}~\bibnamefont {Melia}},\ }\href
  {\doibase 10.23727/CERN-Proceedings-2018-001.65} {\bibfield  {journal}
  {\bibinfo  {journal} {CERN Proc.}\ }\textbf {\bibinfo {volume} {1}},\
  \bibinfo {pages} {65} (\bibinfo {year} {2018})},\ \Eprint
  {http://arxiv.org/abs/1709.07110} {arXiv:1709.07110 [hep-ph]} \BibitemShut
  {NoStop}%
\bibitem [{ATL(2020)}]{ATLAS:2020khf}%
  \BibitemOpen
  \href@noop {} {\  (\bibinfo {year} {2020})}\BibitemShut {NoStop}%
\bibitem [{\citenamefont {Bjorken}\ \emph {et~al.}(1988)\citenamefont
  {Bjorken}, \citenamefont {Ecklund}, \citenamefont {Nelson}, \citenamefont
  {Abashian}, \citenamefont {Church}, \citenamefont {Lu}, \citenamefont {Mo},
  \citenamefont {Nunamaker},\ and\ \citenamefont {Rassmann}}]{Bjorken:1988as}%
  \BibitemOpen
  \bibfield  {author} {\bibinfo {author} {\bibfnamefont {J.}~\bibnamefont
  {Bjorken}}, \bibinfo {author} {\bibfnamefont {S.}~\bibnamefont {Ecklund}},
  \bibinfo {author} {\bibfnamefont {W.}~\bibnamefont {Nelson}}, \bibinfo
  {author} {\bibfnamefont {A.}~\bibnamefont {Abashian}}, \bibinfo {author}
  {\bibfnamefont {C.}~\bibnamefont {Church}}, \bibinfo {author} {\bibfnamefont
  {B.}~\bibnamefont {Lu}}, \bibinfo {author} {\bibfnamefont {L.}~\bibnamefont
  {Mo}}, \bibinfo {author} {\bibfnamefont {T.}~\bibnamefont {Nunamaker}}, \
  and\ \bibinfo {author} {\bibfnamefont {P.}~\bibnamefont {Rassmann}},\ }\href
  {\doibase 10.1103/PhysRevD.38.3375} {\bibfield  {journal} {\bibinfo
  {journal} {Phys. Rev. D}\ }\textbf {\bibinfo {volume} {38}},\ \bibinfo
  {pages} {3375} (\bibinfo {year} {1988})}\BibitemShut {NoStop}%
\bibitem [{\citenamefont {Payez}\ \emph {et~al.}(2015)\citenamefont {Payez},
  \citenamefont {Evoli}, \citenamefont {Fischer}, \citenamefont {Giannotti},
  \citenamefont {Mirizzi},\ and\ \citenamefont {Ringwald}}]{Payez:2014xsa}%
  \BibitemOpen
  \bibfield  {author} {\bibinfo {author} {\bibfnamefont {A.}~\bibnamefont
  {Payez}}, \bibinfo {author} {\bibfnamefont {C.}~\bibnamefont {Evoli}},
  \bibinfo {author} {\bibfnamefont {T.}~\bibnamefont {Fischer}}, \bibinfo
  {author} {\bibfnamefont {M.}~\bibnamefont {Giannotti}}, \bibinfo {author}
  {\bibfnamefont {A.}~\bibnamefont {Mirizzi}}, \ and\ \bibinfo {author}
  {\bibfnamefont {A.}~\bibnamefont {Ringwald}},\ }\href {\doibase
  10.1088/1475-7516/2015/02/006} {\bibfield  {journal} {\bibinfo  {journal}
  {JCAP}\ }\textbf {\bibinfo {volume} {02}},\ \bibinfo {pages} {006} (\bibinfo
  {year} {2015})},\ \Eprint {http://arxiv.org/abs/1410.3747} {arXiv:1410.3747
  [astro-ph.HE]} \BibitemShut {NoStop}%
\bibitem [{\citenamefont {Jaeckel}\ \emph {et~al.}(2018)\citenamefont
  {Jaeckel}, \citenamefont {Malta},\ and\ \citenamefont
  {Redondo}}]{Jaeckel:2017tud}%
  \BibitemOpen
  \bibfield  {author} {\bibinfo {author} {\bibfnamefont {J.}~\bibnamefont
  {Jaeckel}}, \bibinfo {author} {\bibfnamefont {P.}~\bibnamefont {Malta}}, \
  and\ \bibinfo {author} {\bibfnamefont {J.}~\bibnamefont {Redondo}},\ }\href
  {\doibase 10.1103/PhysRevD.98.055032} {\bibfield  {journal} {\bibinfo
  {journal} {Phys. Rev. D}\ }\textbf {\bibinfo {volume} {98}},\ \bibinfo
  {pages} {055032} (\bibinfo {year} {2018})},\ \Eprint
  {http://arxiv.org/abs/1702.02964} {arXiv:1702.02964 [hep-ph]} \BibitemShut
  {NoStop}%
\bibitem [{\citenamefont {Aaboud}\ \emph {et~al.}(2018)\citenamefont {Aaboud}
  \emph {et~al.}}]{Aaboud:2018bun}%
  \BibitemOpen
  \bibfield  {author} {\bibinfo {author} {\bibfnamefont {M.}~\bibnamefont
  {Aaboud}} \emph {et~al.} (\bibinfo {collaboration} {ATLAS}),\ }\href
  {\doibase 10.1103/PhysRevD.98.052008} {\bibfield  {journal} {\bibinfo
  {journal} {Phys. Rev. D}\ }\textbf {\bibinfo {volume} {98}},\ \bibinfo
  {pages} {052008} (\bibinfo {year} {2018})},\ \Eprint
  {http://arxiv.org/abs/1808.02380} {arXiv:1808.02380 [hep-ex]} \BibitemShut
  {NoStop}%
\bibitem [{\citenamefont {Sirunyan}\ \emph {et~al.}(2019)\citenamefont
  {Sirunyan} \emph {et~al.}}]{Sirunyan:2019vgt}%
  \BibitemOpen
  \bibfield  {author} {\bibinfo {author} {\bibfnamefont {A.~M.}\ \bibnamefont
  {Sirunyan}} \emph {et~al.} (\bibinfo {collaboration} {CMS}),\ }\href
  {\doibase 10.1016/j.physletb.2019.134952} {\bibfield  {journal} {\bibinfo
  {journal} {Phys. Lett. B}\ }\textbf {\bibinfo {volume} {798}},\ \bibinfo
  {pages} {134952} (\bibinfo {year} {2019})},\ \Eprint
  {http://arxiv.org/abs/1906.00057} {arXiv:1906.00057 [hep-ex]} \BibitemShut
  {NoStop}%
\bibitem [{\citenamefont {Bellazzini}\ \emph {et~al.}(2013)\citenamefont
  {Bellazzini}, \citenamefont {Csaki}, \citenamefont {Hubisz}, \citenamefont
  {Serra},\ and\ \citenamefont {Terning}}]{Bellazzini:2012vz}%
  \BibitemOpen
  \bibfield  {author} {\bibinfo {author} {\bibfnamefont {B.}~\bibnamefont
  {Bellazzini}}, \bibinfo {author} {\bibfnamefont {C.}~\bibnamefont {Csaki}},
  \bibinfo {author} {\bibfnamefont {J.}~\bibnamefont {Hubisz}}, \bibinfo
  {author} {\bibfnamefont {J.}~\bibnamefont {Serra}}, \ and\ \bibinfo {author}
  {\bibfnamefont {J.}~\bibnamefont {Terning}},\ }\href {\doibase
  10.1140/epjc/s10052-013-2333-x} {\bibfield  {journal} {\bibinfo  {journal}
  {Eur. Phys. J.}\ }\textbf {\bibinfo {volume} {C73}},\ \bibinfo {pages} {2333}
  (\bibinfo {year} {2013})},\ \Eprint {http://arxiv.org/abs/1209.3299}
  {arXiv:1209.3299 [hep-ph]} \BibitemShut {NoStop}%
%%CITATION = ARXIV:1209.3299;%%
\bibitem [{\citenamefont {Sundrum}(2011)}]{Sundrum:2009gv}%
  \BibitemOpen
  \bibfield  {author} {\bibinfo {author} {\bibfnamefont {R.}~\bibnamefont
  {Sundrum}},\ }\href {\doibase 10.1007/JHEP01(2011)062} {\bibfield  {journal}
  {\bibinfo  {journal} {JHEP}\ }\textbf {\bibinfo {volume} {01}},\ \bibinfo
  {pages} {062} (\bibinfo {year} {2011})},\ \Eprint
  {http://arxiv.org/abs/0909.5430} {arXiv:0909.5430 [hep-th]} \BibitemShut
  {NoStop}%
%%CITATION = ARXIV:0909.5430;%%
\bibitem [{\citenamefont {von Harling}\ and\ \citenamefont
  {Servant}(2018)}]{vonHarling:2017yew}%
  \BibitemOpen
  \bibfield  {author} {\bibinfo {author} {\bibfnamefont {B.}~\bibnamefont {von
  Harling}}\ and\ \bibinfo {author} {\bibfnamefont {G.}~\bibnamefont
  {Servant}},\ }\href {\doibase 10.1007/JHEP01(2018)159} {\bibfield  {journal}
  {\bibinfo  {journal} {JHEP}\ }\textbf {\bibinfo {volume} {01}},\ \bibinfo
  {pages} {159} (\bibinfo {year} {2018})},\ \Eprint
  {http://arxiv.org/abs/1711.11554} {arXiv:1711.11554 [hep-ph]} \BibitemShut
  {NoStop}%
%%CITATION = ARXIV:1711.11554;%%
\bibitem [{\citenamefont {Baratella}\ \emph {et~al.}(2019)\citenamefont
  {Baratella}, \citenamefont {Pomarol},\ and\ \citenamefont
  {Rompineve}}]{Baratella:2018pxi}%
  \BibitemOpen
  \bibfield  {author} {\bibinfo {author} {\bibfnamefont {P.}~\bibnamefont
  {Baratella}}, \bibinfo {author} {\bibfnamefont {A.}~\bibnamefont {Pomarol}},
  \ and\ \bibinfo {author} {\bibfnamefont {F.}~\bibnamefont {Rompineve}},\
  }\href {\doibase 10.1007/JHEP03(2019)100} {\bibfield  {journal} {\bibinfo
  {journal} {JHEP}\ }\textbf {\bibinfo {volume} {03}},\ \bibinfo {pages} {100}
  (\bibinfo {year} {2019})},\ \Eprint {http://arxiv.org/abs/1812.06996}
  {arXiv:1812.06996 [hep-ph]} \BibitemShut {NoStop}%
%%CITATION = ARXIV:1812.06996;%%
\bibitem [{\citenamefont {Bloch}\ \emph {et~al.}(2019)\citenamefont {Bloch},
  \citenamefont {Csáki}, \citenamefont {Geller},\ and\ \citenamefont
  {Volansky}}]{Bloch:2019bvc}%
  \BibitemOpen
  \bibfield  {author} {\bibinfo {author} {\bibfnamefont {I.~M.}\ \bibnamefont
  {Bloch}}, \bibinfo {author} {\bibfnamefont {C.}~\bibnamefont {Csáki}},
  \bibinfo {author} {\bibfnamefont {M.}~\bibnamefont {Geller}}, \ and\ \bibinfo
  {author} {\bibfnamefont {T.}~\bibnamefont {Volansky}},\ }\href@noop {} {\
  (\bibinfo {year} {2019})},\ \Eprint {http://arxiv.org/abs/1912.08840}
  {arXiv:1912.08840 [hep-ph]} \BibitemShut {NoStop}%
%%CITATION = ARXIV:1912.08840;%%
\bibitem [{\citenamefont {Konstandin}\ and\ \citenamefont
  {Servant}(2011)}]{Konstandin:2011dr}%
  \BibitemOpen
  \bibfield  {author} {\bibinfo {author} {\bibfnamefont {T.}~\bibnamefont
  {Konstandin}}\ and\ \bibinfo {author} {\bibfnamefont {G.}~\bibnamefont
  {Servant}},\ }\href {\doibase 10.1088/1475-7516/2011/12/009} {\bibfield
  {journal} {\bibinfo  {journal} {JCAP}\ }\textbf {\bibinfo {volume} {12}},\
  \bibinfo {pages} {009} (\bibinfo {year} {2011})},\ \Eprint
  {http://arxiv.org/abs/1104.4791} {arXiv:1104.4791 [hep-ph]} \BibitemShut
  {NoStop}%
\bibitem [{\citenamefont {Agashe}\ \emph {et~al.}(2020)\citenamefont {Agashe},
  \citenamefont {Du}, \citenamefont {Ekhterachian}, \citenamefont {Kumar},\
  and\ \citenamefont {Sundrum}}]{Agashe:2019lhy}%
  \BibitemOpen
  \bibfield  {author} {\bibinfo {author} {\bibfnamefont {K.}~\bibnamefont
  {Agashe}}, \bibinfo {author} {\bibfnamefont {P.}~\bibnamefont {Du}}, \bibinfo
  {author} {\bibfnamefont {M.}~\bibnamefont {Ekhterachian}}, \bibinfo {author}
  {\bibfnamefont {S.}~\bibnamefont {Kumar}}, \ and\ \bibinfo {author}
  {\bibfnamefont {R.}~\bibnamefont {Sundrum}},\ }\href {\doibase
  10.1007/JHEP05(2020)086} {\bibfield  {journal} {\bibinfo  {journal} {JHEP}\
  }\textbf {\bibinfo {volume} {05}},\ \bibinfo {pages} {086} (\bibinfo {year}
  {2020})},\ \Eprint {http://arxiv.org/abs/1910.06238} {arXiv:1910.06238
  [hep-ph]} \BibitemShut {NoStop}%
\bibitem [{\citenamefont {Creminelli}\ \emph {et~al.}(2002)\citenamefont
  {Creminelli}, \citenamefont {Nicolis},\ and\ \citenamefont
  {Rattazzi}}]{Creminelli:2001th}%
  \BibitemOpen
  \bibfield  {author} {\bibinfo {author} {\bibfnamefont {P.}~\bibnamefont
  {Creminelli}}, \bibinfo {author} {\bibfnamefont {A.}~\bibnamefont {Nicolis}},
  \ and\ \bibinfo {author} {\bibfnamefont {R.}~\bibnamefont {Rattazzi}},\
  }\href {\doibase 10.1088/1126-6708/2002/03/051} {\bibfield  {journal}
  {\bibinfo  {journal} {JHEP}\ }\textbf {\bibinfo {volume} {03}},\ \bibinfo
  {pages} {051} (\bibinfo {year} {2002})},\ \Eprint
  {http://arxiv.org/abs/hep-th/0107141} {arXiv:hep-th/0107141} \BibitemShut
  {NoStop}%
\bibitem [{\citenamefont {Bellazzini}\ \emph {et~al.}(2014)\citenamefont
  {Bellazzini}, \citenamefont {Csáki},\ and\ \citenamefont
  {Serra}}]{Bellazzini:2014yua}%
  \BibitemOpen
  \bibfield  {author} {\bibinfo {author} {\bibfnamefont {B.}~\bibnamefont
  {Bellazzini}}, \bibinfo {author} {\bibfnamefont {C.}~\bibnamefont {Csáki}},
  \ and\ \bibinfo {author} {\bibfnamefont {J.}~\bibnamefont {Serra}},\ }\href
  {\doibase 10.1140/epjc/s10052-014-2766-x} {\bibfield  {journal} {\bibinfo
  {journal} {Eur. Phys. J. C}\ }\textbf {\bibinfo {volume} {74}},\ \bibinfo
  {pages} {2766} (\bibinfo {year} {2014})},\ \Eprint
  {http://arxiv.org/abs/1401.2457} {arXiv:1401.2457 [hep-ph]} \BibitemShut
  {NoStop}%
\bibitem [{\citenamefont {Panico}\ and\ \citenamefont
  {Wulzer}(2016)}]{Panico:2015jxa}%
  \BibitemOpen
  \bibfield  {author} {\bibinfo {author} {\bibfnamefont {G.}~\bibnamefont
  {Panico}}\ and\ \bibinfo {author} {\bibfnamefont {A.}~\bibnamefont
  {Wulzer}},\ }\href {\doibase 10.1007/978-3-319-22617-0} {\emph {\bibinfo
  {title} {{The Composite Nambu-Goldstone Higgs}}}},\ Vol.\ \bibinfo {volume}
  {913}\ (\bibinfo  {publisher} {Springer},\ \bibinfo {year} {2016})\ \Eprint
  {http://arxiv.org/abs/1506.01961} {arXiv:1506.01961 [hep-ph]} \BibitemShut
  {NoStop}%
\end{thebibliography}%

\end{document}